\documentclass[aps,prb,showpacs,amsmath,amssymb]{revtex4-1} 
\usepackage{graphicx}

\providecommand{\tabularnewline}{\\}

\begin{document}

\title{Spectrum of $\pi$ electrons in bilayer graphene nanoribbons and
nanotubes: an analytical approach}

\author{J. Ruseckas}

\email{julius.ruseckas@tfai.vu.lt}

\homepage{http://www.itpa.lt/~ruseckas}

\author{G. Juzeli\=unas}

\affiliation{Institute of Theoretical Physics and Astronomy, Vilnius University
A.~Go\v{s}tauto 12, LT-01108 Vilnius, Lithuania}

\author{I. V. Zozoulenko}

\affiliation{Solid State Electronics, ITN, Link\"oping University, 601 74
Nork\"oping,Sweden}

\begin{abstract}
We present an analytical description of $\pi$ electrons of a finite size
bilayer graphene within a framework of the tight-binding model. The bilayered
structures considered here are characterized by a rectangular geometry and have
a finite size in one or both directions with armchair- and zigzag-shaped
edges. We provide an exact analytical description of the spectrum of $\pi$
electrons in the zigzag and armchair bilayer graphene nanoribbons and
nanotubes. We analyze the dispersion relations, the density of states, and the
conductance quantization.
\end{abstract}


\maketitle

\section{introduction}

Since its isolation in 2004, graphene---a single sheet of carbon atoms
arranged in a honeycomb lattice---has attracted an enormous attention because
of its highly unusual electronic and transport properties that are strikingly
different from those of conventional semiconductor-based two-dimensional
electronic systems (for a review see
Refs.~\onlinecite{Review09,advances,reviewDasSarma,reviewPeres}). It has been
immediately realized the significance and the potential impact of this new
material for electronics. This far, it has been demonstrated that the graphene
has the highest carrier mobility at room temperature in comparison to any known
material\cite{Du}. However, graphene is a semimetal with no gap and zero
density of states at the Fermi energy. This makes it difficult to utilize it in
electronic devices such as field effect transistor (FET) requiring a large
on/off current ratio. The energy gap can be opened in a bilayer graphene by
applying a gate voltage between the layers\cite{McCan}. This gate-induced
bandgap was demonstrated by Oostinga \textit{et al.}\cite{Oostinga}, and the
on/off current ratio of around 100 at room temperature for a dual-gate bilayer
graphene FET was reported by the IBM\cite{IBM}.

Another way to introduce the gap is to pattern graphene into
nanoribbons\cite{Wakabayashi99,Son}. The conductance of graphene nanoribbons
(GNRs) with lithographically etched edges indeed revealed the gap in the
transport measurements\cite{Han,Lin}. This gap has been subsequently understood
as the edge-disorder-induced transport gap\cite{Martin,Lewenkopf,Kirczenow}
rather than the intrinsic energy gap expected in ideal GNRs due to the
confinement\cite{Wakabayashi99} or electron interactions and edge
effects\cite{Son} During last years the great progress has been achieved in
fabrication and patterning of the GNRs with ultrasmooth and/or atomically
controlled edges. This includes e.g.\ a controlled formation of edges by Joule
heating\cite{Joule}, unzipping carbon nanotubes to form
nanoribbons\cite{nanotubes,*nanotubes-2}, chemical route to produce nanoribbons
with ultrasmooth edges\cite{chemical} and atomically precise bottom-up
fabrication of GNRs\cite{bottom-up}. All these advances in nanoribbons
fabrication will hopefully enable not before long the electronic measurement in
near-perfect nanoribbons free from the edge or bulk disorder defects.

An important insight into electronic properties of graphene and GNRs can be
obtained from exact analytical approaches. The analytic calculations for the
electronic structure of the GNRs have been reported in
Refs.~\onlinecite{Brey_Fertig,Zheng,OnipkoPRL,Onipko08,Jiang}. The electronic
structure of the bilayer graphene was addressed in
Refs.~\onlinecite{Guinea,Peeters,Wang,Nilsson,Bilayer_edges} where the analytical
results were presented (both exact and perturbative). We are not however aware
of analytical treatment of bilayer GNRs (Note that a numerical study of the
magnetobandstructure of the GNRs was reported in Ref.~\onlinecite{Xu,BGN_Screening},
and the analytical and numerical treatment of the edge states in the
bi- and N-layer graphene and GNRs was presented in
Refs.~\onlinecite{Bilayer_edges,Castro08}). The purpose of the present study is to
provide an exact analytical description of the spectrum of $\pi$ electrons in
the zigzag and armchair bilayer nanoribbons and nanotubes including the
dispersion relations, the density of states, and the conductance quantization.

The paper is organized as follows: In order to illustrate our method, in
Sec.~\ref{sec:graphene} we present known analytical results for a simpler
system, monolayer graphene of the finite size. Subsequently in
Sec.~\ref{sec:bilayer} we derive the main analytical expressions for the energy
spectrum of finite-size structures of bilayer graphene. These expressions are
used in Sec.~\ref{sec:fermi} to analyze the energy spectrum of various bilayer
graphene structures near the Fermi energy. Finally, Sec.~\ref{sec:concl}
summarizes our findings.

\section{Single layer graphene}

\label{sec:graphene}Analytical expressions for the $\pi$ electron spectrum in
GNRs and graphene nanotubes (GNTs), based on tight-binding model, were provided
in Ref.~\onlinecite{Onipko08}. In this section we will rederive the same expressions
in an analytically simpler way. Our method more clearly shows the connection
between solutions for the infinite sheet of graphene and for the finite-size
sheet. In addition, simpler method will allow us to derive later on analytical
expressions of the $\pi$ electron spectrum for more complex systems, bilayer
GNRs and GNTs.

\subsection{Electron spectrum in infinite sheet of graphene}

\begin{figure}
\includegraphics[width=0.65\textwidth]{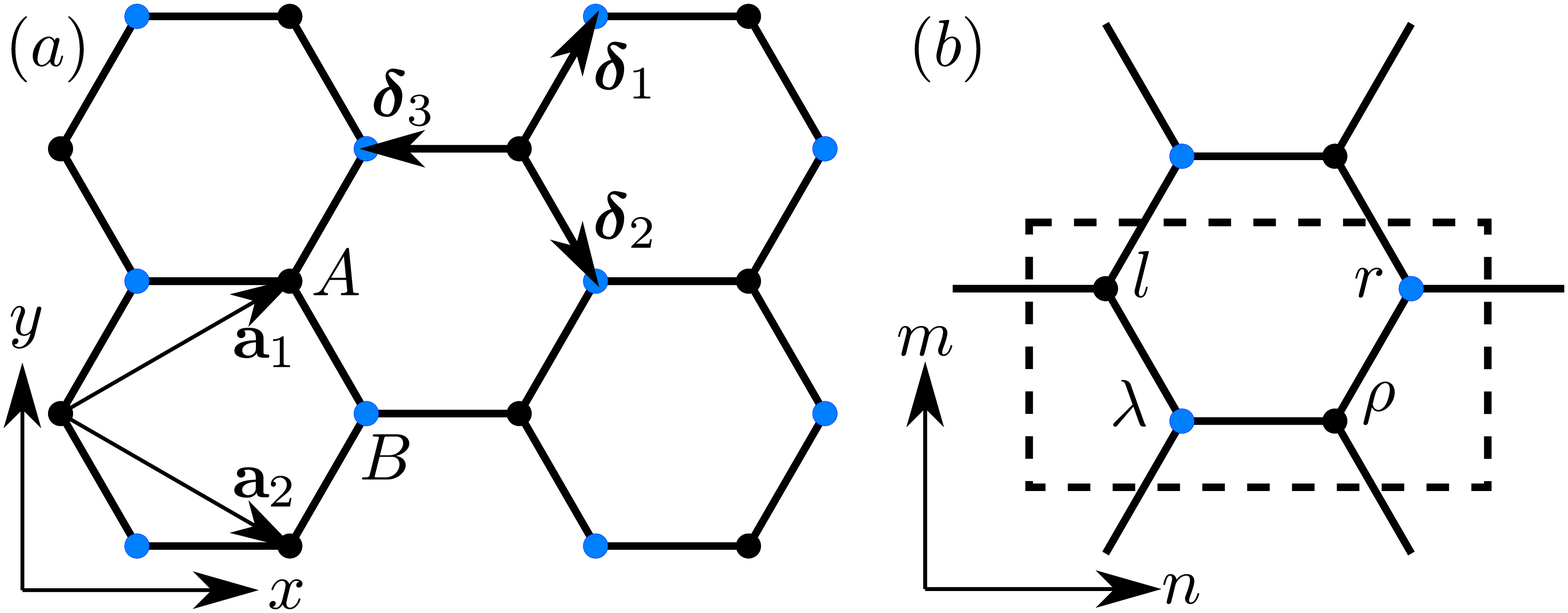}\includegraphics[width=0.25\textwidth]{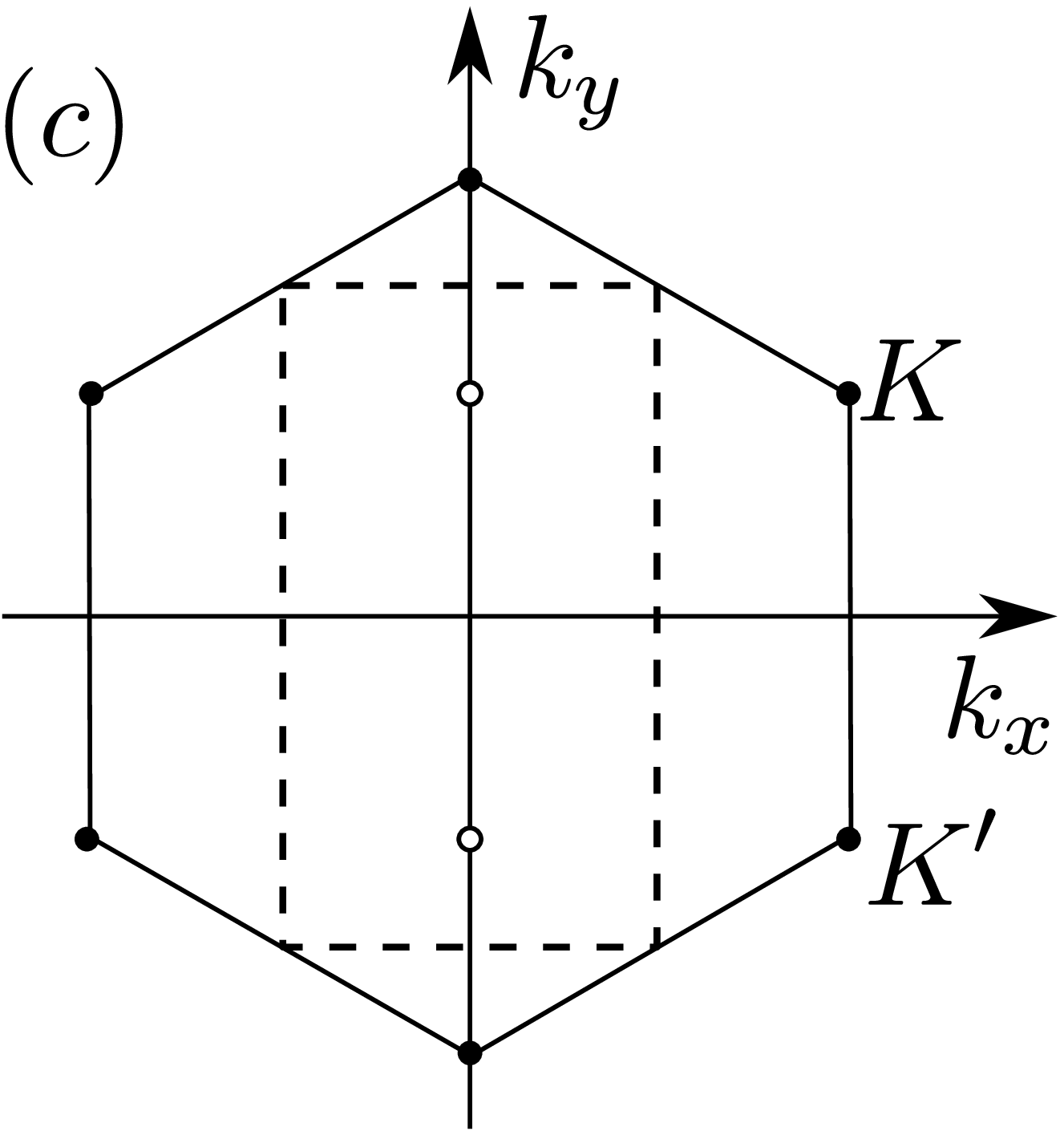}

\caption{(Color online) (a) Honeycomb lattice structure of graphene, made out
of two interpenetrating triangular lattices. $\mathbf{a}_1$ and $\mathbf{a}_2$
are the lattice unit vectors, and $\boldsymbol{\delta}_i$, $i=1,2,3$ are the
nearest-neighbor vectors. (b) Indication of labels of carbon atoms in the
rectangular unit cell. (c) Brillouin zones for hexagonal unit cell (solid
hexagon) and rectangular unit cell (dashed rectangle). The Dirac points are
indicated by solid circles for the hexagonal unit cell and hollow circles for
the rectangular unit cell.}
\label{fig:graphene-1}
\end{figure}

First we will consider $\pi$ electron spectrum in an infinite sheet of
graphene. Hexagonal structure of graphene is shown in
Fig~\ref{fig:graphene-1}a. The structure of the graphene can be viewed as a
hexagonal lattice with a basis of two atoms per unit cell. The Cartesian
components of the lattice vectors $\mathbf{a}_1$ and $\mathbf{a}_2$ are
$a(3/2,\sqrt{3}/2)$ and $a(3/2,-\sqrt{3}/2)$, respectively. Here
$a\approx 1.42\,\AA$ is the carbon-carbon distance\cite{Review09}. The three
nearest-neighbor vectors are given by
$\boldsymbol{\delta}_1=a(1/2,\sqrt{3}/2)$,
$\boldsymbol{\delta}_2=a(1/2,-\sqrt{3}/2)$, and
$\boldsymbol{\delta}_1=a(-1,0)$. The tight-binding Hamiltonian for electrons in
graphene has the form
\begin{equation}
H_{\mathrm{gr}}=-t\sum_{\langle
i,j\rangle}(a_i^{\dag}b_j+b_j^{\dag}a_i)\,,
\label{eq:H-tight}
\end{equation}
where the operators $a_i$ and $b_i$ annihilate an electron on sublattice $A$ at
site $\mathbf{R}_i^A$ and on sublattice $B$ at site $\mathbf{R}_i^B$,
respectively. The parameter $t$ is the nearest-neighbor hopping energy
($t\approx2.8\,\mathrm{eV}$). From now on we will write all energies in the
units of the hoping integral $t$, therefore we will set $t=1$. Let us label the
elementary cells of the lattice with two numbers $p$ and $q$. Then the atoms in
the sublattices $A$ and $B$ are positioned at
$\mathbf{R}_{p,q}^A=p\mathbf{a}_1+q\mathbf{a}_2$ and
$\mathbf{R}_{p,q}^B=\boldsymbol{\delta}_1+p\mathbf{a}_1+q\mathbf{a}_2$,
respectively.

The $\pi$ electron wave function satisfies the Schr\"odinger equation,
\begin{equation}
H\Psi=E\Psi\,.
\label{eq:shroed}
\end{equation}
We search for the eigenvectors of the Hamiltonian (\ref{eq:H-tight}) in the
form of the plane waves (Bloch states) by taking the probability amplitudes to
find an atom in the sites $\mathbf{R}_{p,q}^A$ and $\mathbf{R}_{p,q}^B$ of the
sublattices $A$ and $B$ as
\begin{equation}
\psi_{p,q}^A=c^Ae^{i\mathbf{k}\cdot\mathbf{R}_{p,q}^A}\,,\qquad\psi_{
p,q}^B=c^Be^{i\mathbf{k}\cdot\mathbf{R}_{p,q}^B}\,.
\label{eq:psi-periodic}
\end{equation}
Thus Eq.~(\ref{eq:shroed}) yields the eiganvalue equations for the coefficients
$c^A$ and $c^B$ 
\begin{eqnarray}
-Ec^A & = & c^B\tilde{\phi}(\mathbf{k})\,,
\label{eq:A-1}
\\ -Ec^B & = &
c^A\tilde{\phi}(-\mathbf{k})\,,
\label{eq:B-1}
\end{eqnarray}
where
\begin{equation}
\tilde{\phi}(\mathbf{k})\equiv
e^{i\mathbf{k}\cdot\boldsymbol{\delta}_1}+e^{i\mathbf{k}\cdot\boldsymbol{
\delta}_2}+e^{i\mathbf{k}\cdot\boldsymbol{\delta}_3}\,.
\end{equation}
From Eqs.~(\ref{eq:A-1}) and (\ref{eq:B-1}) we get the eigenenergies and the
corresponding coefficients determining the eigenvectors
\begin{equation}
E(\mathbf{k})=s_1|\tilde{\phi}(\mathbf{k})|\,,\qquad
c^A=-\frac{\tilde{\phi}(\mathbf{k})}{E(\mathbf{k})}\,,\qquad
c^B=1\,,
\label{eq:eigen-1}
\end{equation}
where $s_1=\pm1$. In the anticipation of the rectangular geometry we introduce
dimensionless Cartesian components of the wave vector
\begin{equation}
\kappa=3ak_x\,,\qquad\xi=\sqrt{3}ak_y
\end{equation}
instead of the wave vector components $k_x$ and $k_y$. Then using the
coordinates of the vectors $\boldsymbol{\delta}_j$ we have
\begin{equation}
\tilde{\phi}(\mathbf{k})=e^{-i\frac{\kappa}{3}}+2e^{i\frac{\kappa}{
6}}\cos\left(\frac{\xi}{2}\right)
\end{equation}
and the expression for the eigenenergies becomes\cite{Review09}
\begin{equation}
E(\mathbf{k})=s_1\sqrt{1+4\cos^2\left(\frac{\xi}{2}\right)+4\cos\left(\frac{
\xi}{2}\right)\cos\left(\frac{\kappa}{2}\right)}\,.
\label{eq:energy-1}
\end{equation}

For satisfying boundary conditions it is useful to adopt a larger unit cell
characterized the same geometry as the whole sheet of graphene. Since we are
interested in configurations of the graphene with rectangular geometry, we will
use a rectangular unit cell, as it has been done in Ref.~\onlinecite{Onipko08}. Such
unit cell has four atoms labeled with symbols $l$ , $\lambda$ , $\rho$ , $r$ ,
as it is shown in Fig.~\ref{fig:graphene-1}b. The atoms with labels $l$ and
$\rho$ belong to the sublattice $A$ , the atoms with labels $\lambda$ and $r$
belong to the sublattice $B$. The position of the unit cell is indicated with
two numbers $n$ and $m$. The first Brillouin zone corresponding to the
rectangular unit cell contains the values of the wave vectors $\kappa$, $\xi$
in the intervals $-\pi\leq\kappa<\pi$ , $-\pi\leq\xi<\pi$. We search for the
eigenvectors having the form of plane waves,
\begin{equation}
\psi_{m,n,\alpha}=c_{\alpha}e^{i\xi m+i\kappa n}\,,
\label{eq:periodic-1}
\end{equation}
where $\alpha=l,\rho,\lambda,r$. This solution can be obtained from
Eq.~(\ref{eq:psi-periodic}) using the equalities
\begin{equation}
c_r=c^B\,,\quad
c_{\rho}=c^Ae^{-i\mathbf{k}\cdot\boldsymbol{\delta}_1}\,,\quad
c_{\lambda}=c^Be^{-i\mathbf{k}\cdot\mathbf{a}_1}\,,\quad
c_l=c^Ae^{-i2ak_x}\,.
\label{eq:coeff-hexa-rect}
\end{equation}

The Brillouin zones corresponding to hexagonal and rectangular unit cells are
shown in Fig.~\ref{fig:graphene-1}c. Compared to the area of the Brillouin zone
of the hexagonal unit cell, the area of the Brillouin zone of the rectangular
unit cell is two times smaller. Smaller Brillouin zone leads to the appearance
of additional dispersion branches. Those dispersion branches can be taken into
account by using two values of the wave vector $\kappa$ in
Eqs.~(\ref{eq:energy-1}) and (\ref{eq:eigen-1}), the one with
$-\pi\leq\kappa<\pi$ and another obtained replacing $\kappa$ by $2\pi+\kappa$.
Using Eqs.~(\ref{eq:eigen-1}), (\ref{eq:coeff-hexa-rect}) we obtain the
coefficients of the eigenvectors
\begin{eqnarray}
c_r & = & 1\,,\qquad
c_{\rho}=-e^{-i\frac{\xi}{2}}\frac{\phi(\kappa,\xi)}{E(\kappa,\xi)}\,,
\label{eq:gr-r-rho}
\\ c_l & = &
-s_3e^{-i\frac{\kappa}{2}}\frac{\phi(\kappa,\xi)}{E(\kappa,\xi)}\,,\qquad
c_{\lambda}=s_3e^{-i\frac{1}{2}(\kappa+\xi)}\,,
\label{eq:gr-l-lam}
\end{eqnarray}
where
\begin{equation}
\phi(\kappa,\xi)=s_3e^{-i\frac{\kappa}{2}}+2\cos\left(\frac{\xi}{2}\right)
\end{equation}
and $s_3=\pm1$ indicates the dispersion branches that appear due to the smaller
Brillouin zone. The equation for the energy now becomes
\begin{equation}
E(\kappa,\xi)=s_1\sqrt{1+4\cos^2\left(\frac{\xi}{2}\right)+s_34\cos\left(\frac{
\xi}{2}\right)\cos\left(\frac{\kappa}{2}\right)}\,.
\label{eq:energy}
\end{equation}
This equation has been obtained in\cite{Onipko08}. Zero energy points of the
graphene honeycomb lattice with dispersion relation (\ref{eq:energy-1}) are at
the points $K=(2\pi,2\pi/3)$ and $K'=(2\pi,-2\pi/3)$, where coordinates are
given in $(\kappa,\xi)$ space. $K$ points correspond to the corners of the
first Brillouin zone. Using the Brillouin zone corresponding to the rectangular
unit cell, the zero energy points have coordinates
$\left(0,\pm\frac{2\pi}{3}\right)$ and the number of these points is only two,
as it is shown in Fig.~\ref{fig:graphene-1}c.

\begin{figure}
\includegraphics[width=0.45\textwidth]{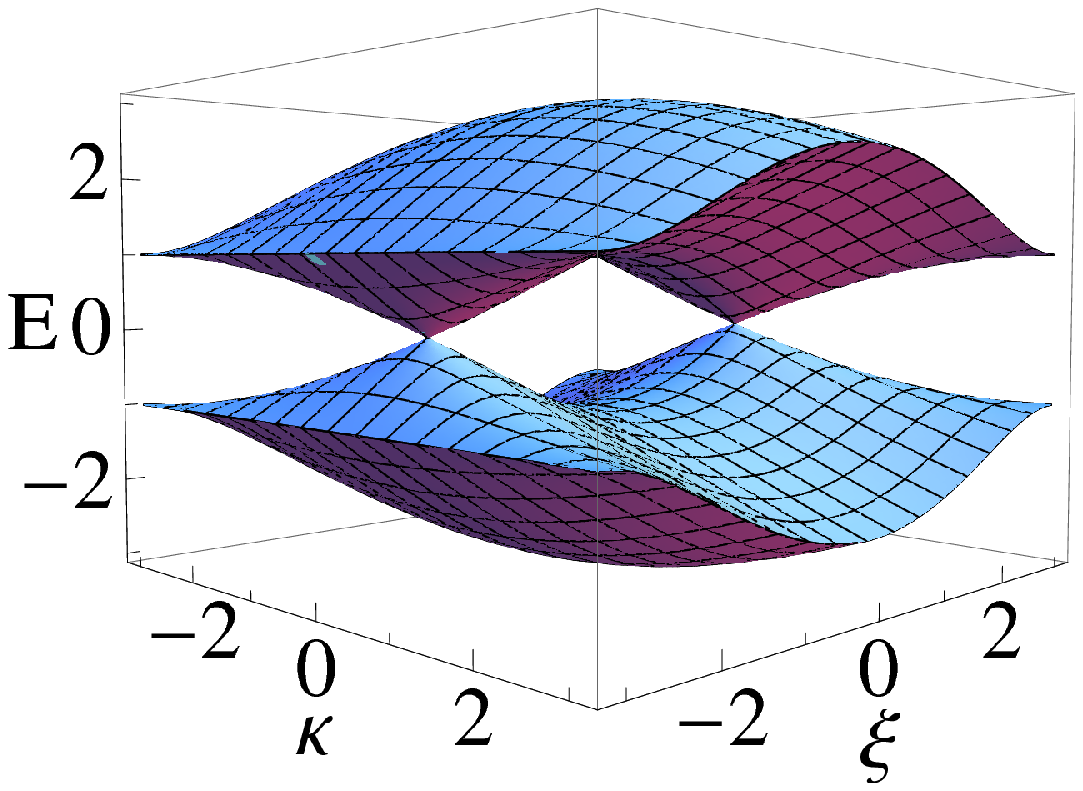}\includegraphics[width=0.5\textwidth]{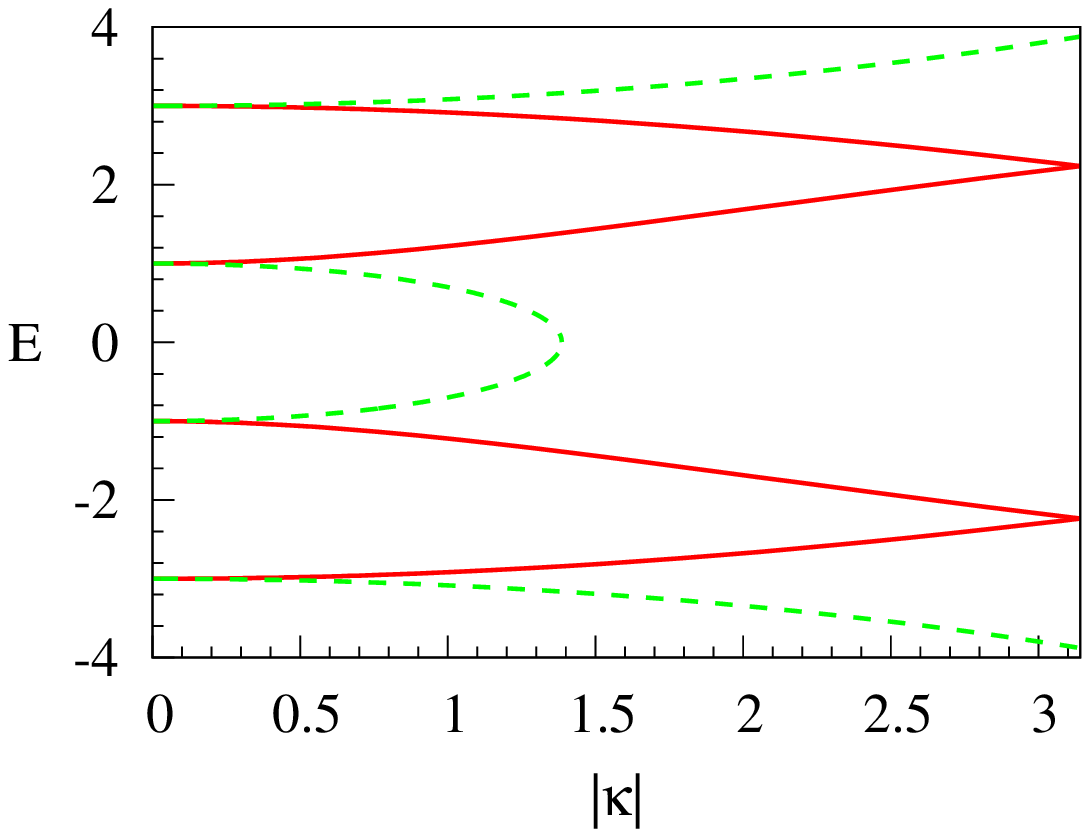}

\caption{(Color online) Left: dispersion branches of
graphene for rectangular unit cell, calculated according to
Eq.~(\ref{eq:energy}). Right: dispersion branches for $\xi=0$, showing
propagating solutions (red solid) and evanescent solutions (green dashed).}
\label{fig:disp-gr}
\end{figure}

Since we will consider finite-size graphene sheets, evanescent solutions become
important. Solution exponentially decreasing or increasing in the $x$-direction
can be obtained by taking $\kappa=i|\kappa|$ in Eqs.~(\ref{eq:gr-r-rho}),
(\ref{eq:gr-l-lam}) and (\ref{eq:energy}), whereas solution exponentially
decreasing or increasing in the $y$-direction can be obtained by taking
$\xi=i|\xi|$. The dependency of the energy on $\kappa$ when $\xi=0$ is shown in
Fig.~\ref{fig:disp-gr}. We see that the branches with real and imaginary
$\kappa$ do not intersect at $|\kappa|>0$.

\subsection{Electron spectrum in various single layer graphene structures}

From the boundary conditions we get restrictions on the possible values of the
wave vectors $\kappa$, $\xi$. We will consider the structures of graphene that
have a set of $N$ rectangular unit cells in the $x$ (armchair) direction and a
set of $\mathcal{N}+1/2$ rectangular unit cells in the $y$ (zigzag) direction,
so that there are $\mathcal{N}$ hexagons along the $y$ axis. Note that
rectangular unit cell shown in Fig.~\ref{fig:graphene-1}b extends over the whole
hexagon in the $y$ direction, whereas it extends over more that one hexagon in
$x$ direction.

Using periodic boundary condition, corresponding to the graphene torus, we get
that the possible values of the wave vectors $\kappa$, $\xi$ are
\begin{eqnarray}
\xi_j & = &\frac{2\pi}{\mathcal{N}}j\,,\qquad j=-\left\lfloor
\frac{\mathcal{N}}{2}\right\rfloor ,-\left\lfloor
\frac{\mathcal{N}}{2}\right\rfloor +1,\ldots,\left\lfloor
\frac{\mathcal{N}-1}{2}\right\rfloor
\label{eq:ksi-periodic}
\\\kappa_{\nu}& =
&\frac{2\pi}{N}\nu\,,\qquad\nu=-\left\lfloor\frac{N}{2}\right\rfloor
,-\left\lfloor\frac{N}{2}\right\rfloor +1,\ldots,\left\lfloor
\frac{N-1}{2}\right\rfloor
\label{eq:kappa-periodic}
\end{eqnarray}
Here $\left\lfloor\cdot\right\rfloor$ denotes the integer part of a number.
Thus the spectrum of graphene torus is given by Eq.~(\ref{eq:energy}) replacing
$\kappa$ and $\xi$ by $\kappa_{\nu}$ and $\xi_j$.

For graphene armchair nanotubes one has the periodic boundary condition in the
$x$ direction and the requirement
$\psi_{0,n,r}=\psi_{0,n,l}=\psi_{\mathcal{N}+1,n,l}=\psi_{\mathcal{N}+1,n,r}=0$
for the $y$ direction. Since the energy (\ref{eq:energy}), does not depend on
the sign of wave vector $\xi$, we will search for the eigenvectors of the
Hamiltonian (\ref{eq:H-tight}) as a superposition of periodic solutions
Eq.~(\ref{eq:periodic-1}) with $\xi$ and $-\xi$,
\begin{equation}
\psi_{m,n,\alpha}=ac_{\alpha}(\xi,\kappa_{\nu})e^{i\xi
m+i\kappa_{\nu}n}+bc_{\alpha}(-\xi,\kappa_{\nu})e^{-i\xi m+i\kappa_{\nu}n}\,,
\end{equation}
where $\kappa_{\nu}$ is given by Eq.~(\ref{eq:kappa-periodic}) and $\xi$ needs
to be determined. From the boundary conditions we get a system of two equations
for the coefficients $a$ and $b$
\begin{eqnarray}
ac_{r,l}(\xi,\kappa_{\nu})+bc_{r,l}(-\xi,\kappa_{\nu})& = & 0\,,\\
ae^{i\xi(\mathcal{N}+1)}c_{r,l}(\xi,\kappa_{\nu})+be^{-i\xi(\mathcal{N}+1)}c_{
r,l}(-\xi,\kappa_{\nu})& = & 0\,.
\end{eqnarray}
This system of equations has non-zero solutions only when the determinant is
zero. From Eqs.~(\ref{eq:gr-r-rho}), (\ref{eq:gr-l-lam}) it follows that the
coefficients $c_{r,l}(\xi,\kappa)$ do not depend on the sign of $\xi$ and we
get the condition $\sin(\xi(\mathcal{N}+1))=0$ or
\begin{equation}
\xi=\frac{\pi j}{\mathcal{N}+1}\,,\qquad
j=1,\ldots,\mathcal{N}
\label{eq:cond-ksi}
\end{equation}
Additionally, there are two $N$-fold degenerate levels corresponding to
$\xi=\pi$ with energies $E=\pm1$. The states of those levels have zero wave
function amplitudes at the $l$ and $r$ sites.

For graphene zigzag nanotubes one has the periodic boundary condition in the
$y$ direction and the condition $\psi_{m,0,r}=\psi_{m,N+1,l}=0$ for the $x$
direction. Similarly as for the armchair nanotubes, the energy
(\ref{eq:energy}), does not depend on the sign of wave vector $\kappa$, and we
search for the eigenvectors of the Hamiltonian (\ref{eq:H-tight}) as a
superposition of periodic solutions Eq.~(\ref{eq:periodic-1}) with $\kappa$ and
$-\kappa$, 
\begin{equation}
\psi_{m,n,\alpha}=ac_{\alpha}(\xi_j,\kappa)e^{i\xi_jm+i\kappa
n}+bc_{\alpha}(\xi_j,-\kappa)e^{i\xi_jm-i\kappa n}\,,
\end{equation}
where $\xi_j$ is given by Eq.~(\ref{eq:ksi-periodic}) and $\kappa$ needs to be
determined. From the boundary conditions we get a system of two equations for
the coefficients $a$ and $b$
\begin{eqnarray}
ac_r(\xi_j,\kappa)+bc_r(\xi_j,-\kappa)& = & 0\,,\\
ac_l(\xi_j,\kappa)e^{i\kappa(N+1)}+bc_l(\xi_j,-\kappa)e^{-i\kappa(N+1)}
& = & 0\,.
\end{eqnarray}
Using Eqs.~(\ref{eq:gr-r-rho}), (\ref{eq:gr-l-lam}) we obtain that non-zero
solutions are possible when
\begin{equation}
\frac{\sin(\kappa
N)}{\sin\left(\kappa\left(N+\frac{1}{2}\right)\right)}=-s_32\cos\left(\frac{
\xi_j}{2}\right)\,.
\label{eq:cond-kappa}
\end{equation}
The possible values of wave vector $\kappa$ should obey this equation. The same
condition has been obtained in Ref.~\onlinecite{Onipko08}. Equation
(\ref{eq:cond-kappa}) allows for the imaginary values of wave vector $\kappa$.
The imaginary values appear when $\xi^c<|\xi_j|<\pi$ and $s_3=-1$, where the
critical value $\xi^c=2\arccos\bigl(N/(2N+1)\bigr)$ of the wave vector $\xi$ is
obtained from Eq.~(\ref{eq:cond-kappa}) setting $\kappa=0$. In the limit
$N\rightarrow\infty$ from the condition (\ref{eq:cond-kappa}) with imaginary
$\kappa$ and Eq.~(\ref{eq:energy}) follows that $E=0$: edge states near zigzag
edges in the semi-infinite system have zero energy.

For $N\times\mathcal{N}$ sheet of graphene open boundary conditions in the $y$
direction are the same as for armchair nanotubes and in the $x$ direction are
the same as for zigzag nanotubes. Since the resulting conditions for the wave
vectors $\kappa$, $\xi$ are not coupled, the eigenvector of the Hamiltonian
(\ref{eq:H-tight}) is a superposition of four periodic solutions having all
possible combinations of the signs of $\kappa$ and $\xi$ and the possible
values of the wave vectors are given by Eqs.~(\ref{eq:cond-ksi}) and
(\ref{eq:cond-kappa}). In addition there are two $N$-fold degenerate levels
corresponding to $\xi=\pi$ with energies $E=\pm1$.

\section{Bilayer graphene}

\label{sec:bilayer}Now we will consider the spectrum of $\pi$ electrons in
bilayer graphene. The tight-binding Hamiltonian for electrons in bilayer
graphene has the form
\begin{eqnarray}
H_{\mathrm{bi}} & = &
V\sum_j(a_{j,2}^{\dag}a_{j,2}+b_{j,2}^{\dag}b_{j,2}-a_{j,1}^{\dag}a_{j,1}-b_{
j,1}^{\dag}b_{j,1})\nonumber\\ & & -t\sum_{\langle
i,j\rangle,p}(a_{i,p}^{\dag}b_{j,p}+b_{j,p}^{\dag}a_{i,p})-t_{\bot}\sum_j(a_{
j,1}^{\dag}a_{j,2}+a_{j,2}^{\dag}a_{j,1})\,,
\label{eq:H-bi-tight}
\end{eqnarray}
where the operators $a_{i,p}$ and $b_{i,p}$ annihilate an electron on
sublattice $A_p$ at site $\mathbf{R}_i^{A_p}$ and on sublattice $B_p$ at site
$\mathbf{R}_i^{B_p}$, respectively. The index $p=1,2$ numbers the layers in the
bilayer system. In the Hamiltonian (\ref{eq:H-bi-tight}) we neglected the terms
corresponding to the hopping between atom $B_1$ and atom $B_2$, with the
hopping energy $\gamma_3$, and the terms corresponding to the hopping between
atom $A_1$ ($A_2$) and and atom $B_2$($B_1$) with the hopping energy
$\gamma_4$. Neglect of those hopping terms leads to the minimal model of
bilayer graphene\cite{Nilsson}. The parameter $t_{\bot}$
($t_{\bot}\approx0.4\,\mathrm{eV}$) is the hopping energy between atom $A_1$
and atom $A_2$ while $V$ is half the shift in the electrochemical potential
between the two layers. Similarly as for the monolayer graphene, we will
express all the energies in the units of $t$.

\subsection{Electron spectrum in infinite sheet of bilayer graphene}

We will proceed similarly as in the previous Section and will analyze an
infinite system at first. The atoms in the sublattices $A_1$ and $A_2$ are
positioned at $\mathbf{R}_{p,q}^{A_{1,2}}=p\mathbf{a}_1+q\mathbf{a}_2$, in the
sublattice $B_1$ the atoms are positioned at
$\mathbf{R}_{p,q}^{B_1}=\boldsymbol{\delta}_1+p\mathbf{a}_1+q\mathbf{a}_2$ and
in the sublattice $B_2$ the atoms are positioned at
$\mathbf{R}_{p,q}^{B_2}=-\boldsymbol{\delta}_1+p\mathbf{a}_1+q\mathbf{a}_2$. We
search for the eigenvectors of the Hamiltonian (\ref{eq:H-bi-tight}) in the
form of the plane waves. The probability amplitudes to find an atom in the
sites $\mathbf{R}_{p,q}^{A_{1,2}}$ and $\mathbf{R}_{p,q}^{B_{1,2}}$ of the
sublattices $A_j$ and $B_j$ are
\begin{equation}
\psi_{p,q}^{A_{1,2}}=c^{A_{1,2}}e^{i\mathbf{k}\cdot\mathbf{R}_{p,q}^{A_{
1,2}}}\,,\qquad\psi_{p,q}^{B_{1,2}}=c^{B_{1,2}}e^{i\mathbf{k}\cdot\mathbf{R}_{
p,q}^{B_{1,2}}}\,.
\label{eq:psi-periodic-bi}
\end{equation}
The coefficients $c^{A_p}$ and $c^{B_p}$ obey the eigenvalue equations
\begin{eqnarray}
-Ec^{A_1} & = & Vc^{A_1}+c^{B_1}\tilde{\phi}(\mathbf{k})+\gamma
c^{A_2}\,,\\ -Ec^{B_1} & = &
Vc^{B_1}+c^{A_1}\tilde{\phi}(-\mathbf{k})\,,\\ -Ec^{A_2} & = &
-Vc^{A_2}+c^{B_2}\tilde{\phi}(-\mathbf{k})+\gamma c^{A_1}\,,\\
-Ec^{B_2} & = & -Vc^{B_2}+c^{A_2}\tilde{\phi}(\mathbf{k})\,.
\end{eqnarray}
Here energy $E$, potential $V$ and interaction between layers $\gamma\equiv
t_{\bot}/t$ are in the units of the hoping integral $t$. Using the
nearest-neighbor hopping energy $t\approx2.8\,\mathrm{eV}$ and the hopping
energy between two layers $t_{\bot}\approx0.4\,\mathrm{eV}$ one gets
$\gamma\approx0.14$. When $V=0$, the $\pi$ electron spectrum is determined by
the equation
\begin{equation}
E(\mathbf{k})=s_1\left(s_2\frac{\gamma}{2}+\sqrt{\frac{\gamma^2}{4}+|\tilde{
\phi}(\mathbf{k})|^2}\right)\,,
\end{equation}
where $s_1,s_2=\pm1$. The coefficients of the eigenvector are
\begin{eqnarray}
c^{A_1} & = & -\frac{E(\mathbf{k})}{\tilde{\phi}(-\mathbf{k})}\,,\qquad
c^{B_1}=1\,,
\label{eq:eigen-bi}
\\ c^{A_2} & = &
s_1s_2\frac{E(\mathbf{k})}{\tilde{\phi}(-\mathbf{k})}\,,\qquad
c^{B_2}=-s_1s_2\frac{\tilde{\phi}(\mathbf{k})}{\tilde{\phi}(-\mathbf{
k})}\,.\nonumber
\end{eqnarray}
When $V\neq0$ the spectrum is
\begin{equation}
E(\mathbf{k})=s_1\sqrt{\frac{\gamma^2}{2}+V^2+|\tilde{\phi}(\mathbf{k})|^2
+s_2\sqrt{\frac{\gamma^4}{4}+|\tilde{\phi}(\mathbf{k})|^2(4V^2+\gamma^2)}}
\end{equation}
and the coefficients of the eigenvector are
\begin{eqnarray}
c^{A_1} & = & -\frac{E(\mathbf{k})+V}{\tilde{\phi}(-\mathbf{k})}\,,\qquad
c^{B_1}=1\,,
\label{eq:eigen-bi-V}
\\ c^{A_2} & = &
\frac{E(\mathbf{k})-V}{\tilde{\phi}(-\mathbf{k})}f(\mathbf{k})\,,\qquad
c^{B_2}=-\frac{\tilde{\phi}(\mathbf{k})}{\tilde{\phi}(-\mathbf{k})}f(\mathbf{
k})\,,\nonumber
\end{eqnarray}
where the function
\begin{equation}
f(\mathbf{k})=\frac{(E(\mathbf{k})+V)^2-|\tilde{\phi}(\mathbf{k})|^2}{
\gamma(E(\mathbf{k})-V)}
\end{equation}
describes the contribution of the second sheet of graphene to the eigenvector.

\begin{figure}
\includegraphics[width=0.9\textwidth]{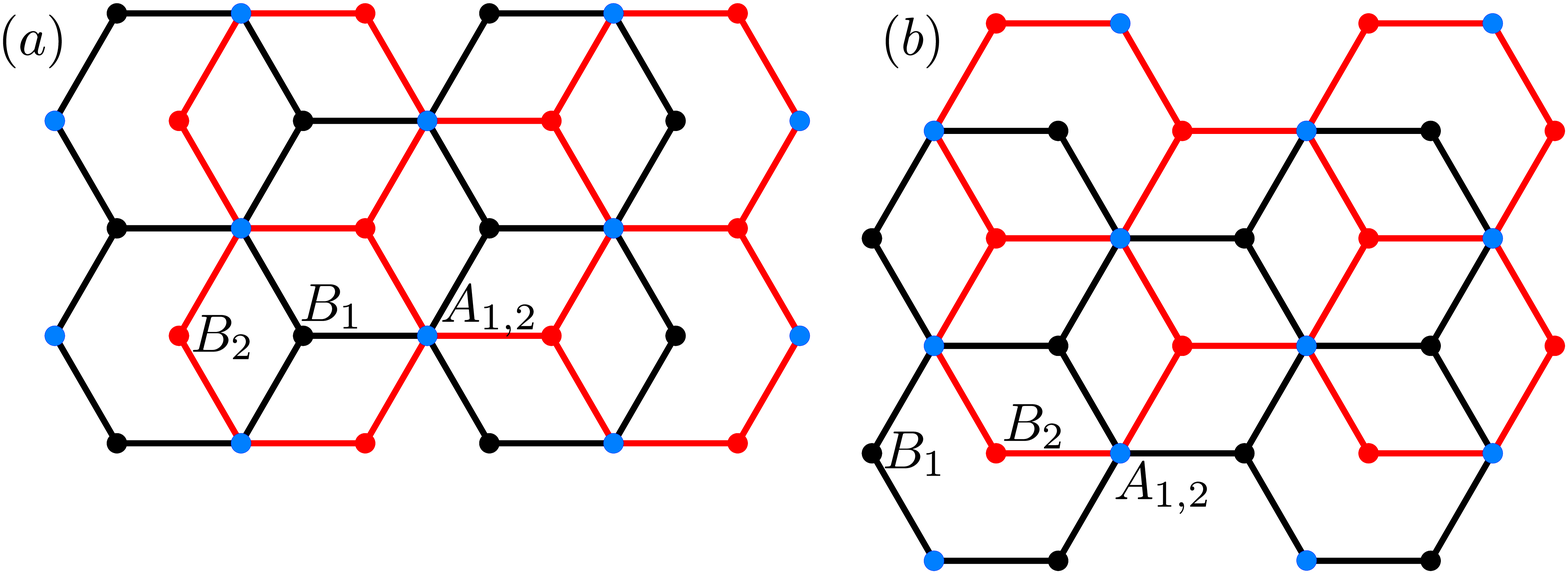}\newline\includegraphics[width=0.7\textwidth]{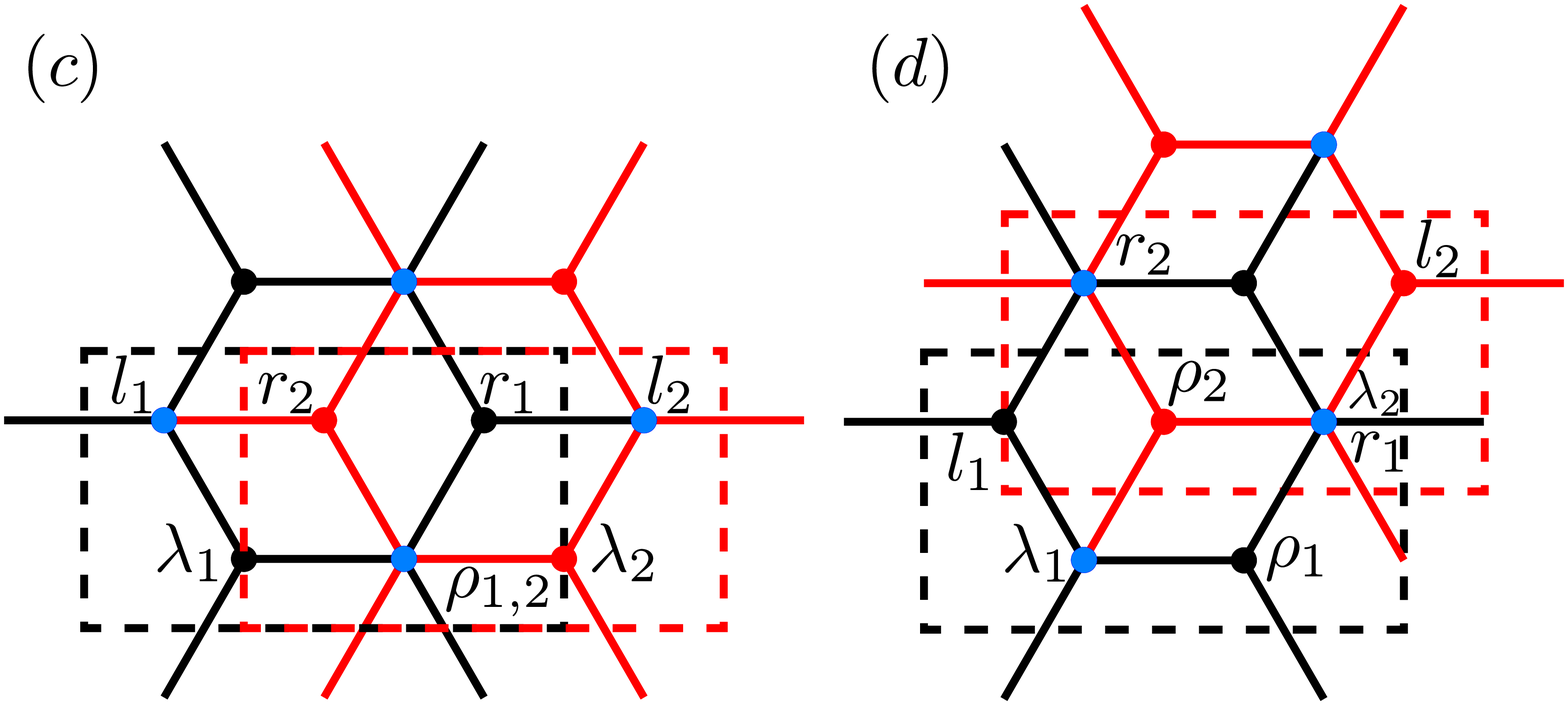}

\caption{(Color online) Upper part: sublatices $A_1$ , $A_2$ , $B_1$ , $B_2$ on
bilayer graphene in AB-$\alpha$ stacking (a) and AB-$\beta$ stacking (b). Lower
part: indication of labels of carbon atoms used in the description of the $\pi$
electron spectrum for the bilayer graphene with AB-$\alpha$ stacking (c) and
AB-$\beta$ stacking (d).}
\label{fig:alpha-beta}
\end{figure}

Finite-size bilayer graphene sheets can be in AB-$\alpha$ or AB-$\beta$
stacking, as is shown in Fig.~\ref{fig:alpha-beta}a,b. Similarly as for
graphene monolayer, we will use rectangular unit cells, one shifted with
respect to the other, in each layer of bilayer graphene. However, the position
of rectangular cells are different for different stacking types. Rectangular
unit cells have eight atoms with labels $l_1$ , $\lambda_1$ , $\rho_1$ , $r_1$
and $l_2$ , $\lambda_2$ , $\rho_2$ , $r_2$ , as is shown in
Fig.~\ref{fig:alpha-beta}c,d. For the AB-$\alpha$ stacking the atoms with
labels $l_1$, $\rho_1$ belong to the sublattice $A_1$ , atoms $\lambda_1$,
$r_1$ to the sublattice $B_1$, atoms $l_2$, $\rho_2$ to the sublattice $A_2$
and atoms $\lambda_2$, $r_2$ to the sublattice $B_2$. For the AB-$\beta$
stacking the atoms with labels $l_1$, $\rho_1$ belong to the sublattice $B_1$ ,
atoms $\lambda_1$, $r_1$ to the sublattice $A_1$, atoms $l_2$, $\rho_2$ to the
sublattice $B_2$ and atoms $\lambda_2$, $r_2$ to the sublattice $A_2$.

We search for the solutions of the form
\begin{equation}
\psi_{m,n,\alpha_p}=c_{\alpha_p}e^{i\xi m+i\kappa n}
\end{equation}
where $\alpha=l,\rho,\lambda,r$ is the label of atoms and $p=1,2$ is the number
of the layer. For the AB-$\alpha$ stacking this solution can be obtained from
Eq.~(\ref{eq:psi-periodic-bi}) using the equalities
\begin{eqnarray}
c_{r_1} & = & c^{B_1}\,,\quad
c_{\rho_1}=c^{A_1}e^{-i\mathbf{k}\cdot\boldsymbol{\delta}_1}\,,\quad
c_{\lambda_1}=c^{B_1}e^{-i\mathbf{k}\cdot\mathbf{a}_1}\,,\quad
c_{l_1}=c^{A_1}e^{-i2ak_x}\,,\\ c_{r_2} & = &
c^{B_2}e^{-iak_x}\,,\quad
c_{\rho_2}=c^{A_2}e^{-i\mathbf{k}\cdot\boldsymbol{\delta}_1}\,,\quad
c_{\lambda_2}=c^{B_2}e^{i\mathbf{k}\cdot\boldsymbol{\delta}_2}\,,\quad
c_{l_2}=c^{A_2}e^{iak_x}
\end{eqnarray}
whereas for the AB-$\beta$ stacking the coefficients are
\begin{eqnarray}
c_{r_1} & = & (c^{A_1})^*\,,\quad
c_{\rho_1}=(c^{B_1})^*e^{-i\mathbf{k}\cdot\boldsymbol{\delta}_1}\,,\quad c_{
\lambda_1}=(c^{A_1})^*e^{-i\mathbf{k}\cdot\mathbf{a}_1}\,,\quad
c_{l_1}=(c^{B_1})^*e^{-i2ak_x}\,,\\ c_{r_2} & = &
(c^{A_2})^*e^{-i\mathbf{k}\cdot\mathbf{a}_2}\,,\quad
c_{\rho_2}=(c^{B_2})^*e^{-iak_x}\,,\quad
c_{\lambda_2}=(c^{A_2})^*\,,\quad
c_{l_2}=(c^{B_2})^*e^{i\mathbf{k}\cdot\boldsymbol{\delta}_1}\,.
\end{eqnarray}

Similarly as for monolayer graphene, to take into account the smaller Brillouin
zone we need two dispersion branches: one with $\kappa$ and one with
$2\pi+\kappa$ . Using Eq.~(\ref{eq:eigen-bi}) or Eq.~(\ref{eq:eigen-bi-V}) we
obtain the coefficients of the eigenvectors. The expressions for the
coefficients are presented in Appendix \ref{appendixA}. The expression for the
energy becomes
\begin{equation}
E(\kappa,\xi)=s_1\sqrt{\frac{\gamma^2}{2}+V^2+|\phi(\kappa,\xi)|^2+s_2\sqrt{
\frac{\gamma^4}{4}+|\phi(\kappa,\xi)|^2(4V^2+\gamma^2)}}
\label{eq:energy-bi-V}
\end{equation}
which reduces to
\begin{equation}
E(\kappa,\xi)=s_1\left(s_2\frac{\gamma}{2}+\sqrt{\frac{\gamma^2}{4}
+|\phi(\kappa,\xi)|^2}\right)
\label{eq:energy-bi}
\end{equation}
for $V=0$. Here
\begin{equation}
|\phi(\kappa,\xi)|^2=1+4\cos^2\left(\frac{\xi}{2}\right)+s_34\cos\left(\frac{
\xi}{2}\right)\cos\left(\frac{\kappa}{2}\right)
\label{eq:phi2}
\end{equation}
and $s_3=\pm1$ indicates the dispersion branches that appear due to the smaller
Brillouin zone.

\begin{figure}
\includegraphics[width=0.4\textwidth]{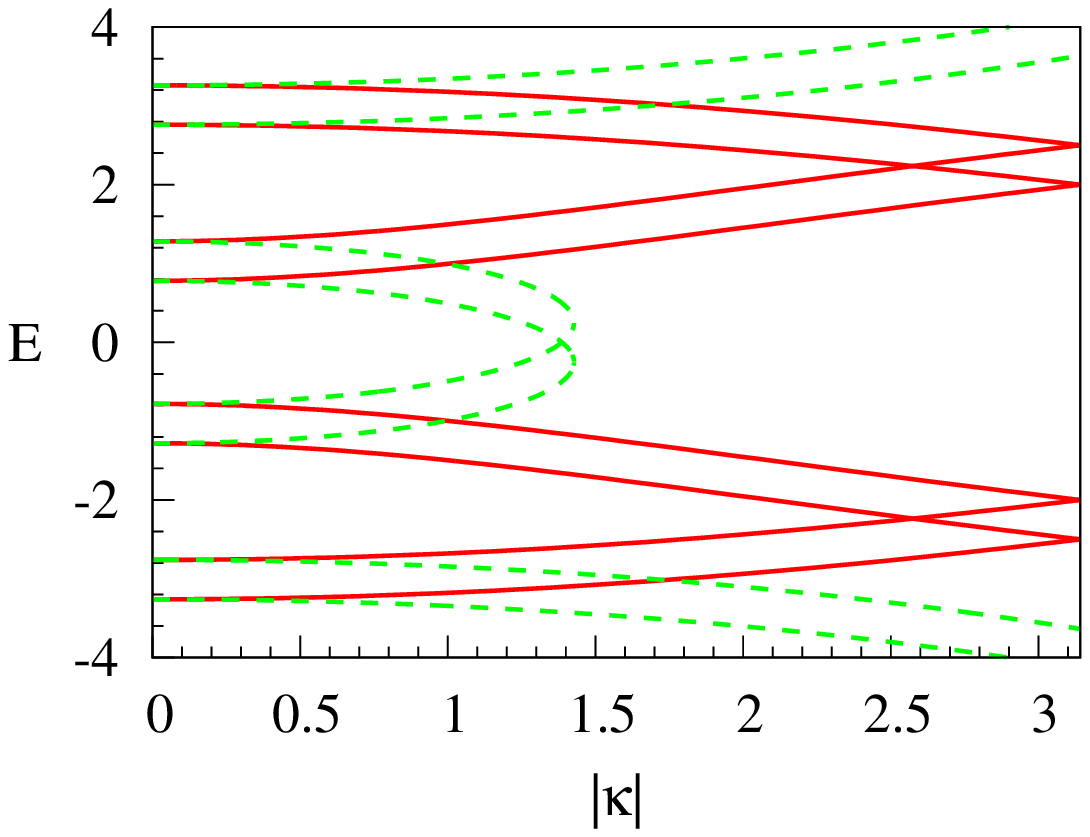}\quad{}\includegraphics[width=0.4\textwidth]{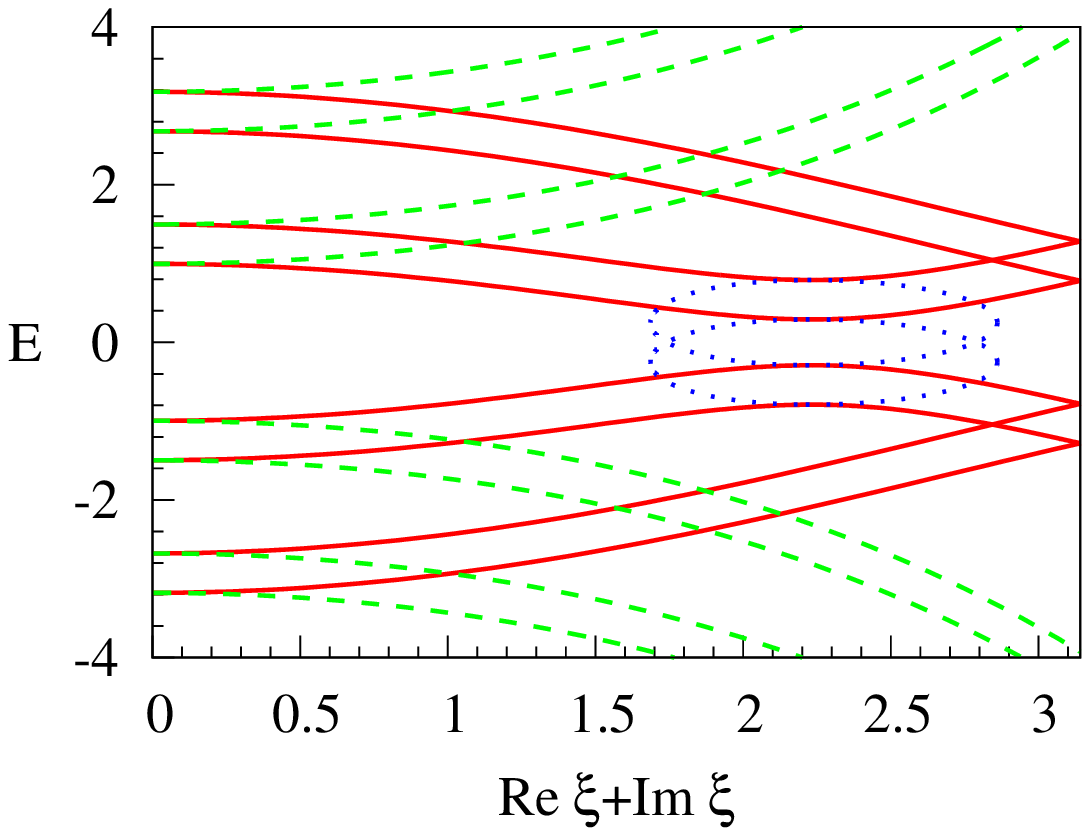}

\caption{(Color online) Dispersion branches of bilayer graphene: dependency of
the energy on the wave vector $\kappa$ when the wave vector $\xi$ is constant
($\xi=0$) (left) and on the wave vector $\xi$ when the wave vector $\kappa$ is
constant ($\kappa=1.0$) (right). Propagating solutions are shown with red solid
line, evanescent solutions with green dashed line, and evanescent oscillating
solutions with complex value of the wave vector $\xi$ are shown with blue
dotted line. In order to show the structure of the dispersion branches more
clearly, the value of the parameter $\gamma$ is set sufficiently large,
$\gamma=0.5$.}
\label{fig:disp-bi}
\end{figure}

In addition to the propagating waves, for finite-size bilayer graphene sheets
evanescent solutions become important. Solution exponentially decreasing or
increasing in the $x$-direction can be obtained by taking $\kappa=i|\kappa|$.
Solution exponentially decreasing or increasing in the $y$-direction can be
obtained by taking $\xi=i|\xi|$. In addition to the purely imaginary $\xi$
there are solutions, corresponding to $s_3=-1$, having complex values of $\xi$.
The dependency of the energy on the wave vector $\kappa$ when the wave vector
$\xi$ is constant and on the wave vector $\xi$ when the wave vector $\kappa$ is
constant is shown in Fig.~\ref{fig:disp-bi}. We see that now, in contrast to
the graphene monolayer, the branches with real and imaginary $\kappa$ can have
the same energy.

\subsection{Electron spectrum in various bilayer graphene structures}

We will consider the structures of bilayer graphene that have a set of $N$
rectangular unit cells in the $x$ (armchair) direction and a set of
$\mathcal{N}+1/2$ rectangular unit cells in the $y$ (zigzag) direction, so that
there are $\mathcal{N}$ hexagons along the $y$ axis. Note that the rectangular
unit cells shown in Figs.~\ref{fig:alpha-beta}c and \ref{fig:alpha-beta}d extend
over the whole hexagon in the $y$ direction, whereas they extend over more that
one hexagon in $x$ direction. In principle, in the case of bilayer graphene
nanotubes the numbers $N$ or $\mathcal{N}$ in for the inner and outer cylinders
are different. However, for simplicity we will consider them as the same, which
is a good approximation for sufficiently large tubes when $N\rightarrow\infty$
or $\mathcal{N}\rightarrow\infty$.

Similarly as for graphene monolayer, from the boundary conditions we get
restrictions on the possible values of the wave vectors $\kappa$, $\xi$. Using
periodic boundary condition, corresponding to the bilayer graphene torus, we
get that the possible values of the wave vectors $\kappa$, $\xi$ are given by
Eqs.~(\ref{eq:ksi-periodic}), (\ref{eq:kappa-periodic}).

For bilayer graphene armchair nanotubes one has the periodic boundary condition
in the $x$ direction and the condition
\begin{equation}
\psi_{0,n,r_p}=\psi_{0,n,l_p}=\psi_{\mathcal{N}+1,n,l_p}=\psi_{\mathcal{N}
+1,n,r_p}=0
\label{eq:b-armchair}
\end{equation}
for the $y$ direction. Here $p=1,2$ is the number of the layer. This condition
is the same for both the AB-$\alpha$ and AB-$\beta$ stackings. For bilayer
graphene with AB-$\alpha$ stacking the coefficients $c_{r_p,l_p}(\xi,\kappa)$
do not depend on the sign of $\xi$ and we get the same conditions
(\ref{eq:kappa-periodic}), (\ref{eq:cond-ksi}) for the wave vectors $\kappa$,
$\xi$, as for the monolayer graphene armchair tubes.

For bilayer graphene with AB-$\beta$ stacking the coefficients
$c_{r_p,l_p}(\xi,\kappa)$ depend on the sign of $\xi$, and condition for the
possible values of the wave vector $\xi$ is much more complicated. There are
eight boundary conditions in the $y$ direction. In bilayer graphene there are
four eigenstates with different wave vectors along $y$ direction, $\xi^{(1)}$,
$\xi^{(2)}$, $\xi^{(3)}$ and $\xi^{(4)}$, having the same energy:
$E(\kappa,\xi^{(1)})=E(\kappa,\xi^{(2)})=E(\kappa,\xi^{(3)})=E(\kappa,\xi^{
(4)})$, as is evident from Fig.~\ref{fig:disp-bi}. Two or four of the wave
vectors $\xi^{(1)}$, $\xi^{(2)}$, $\xi^{(3)}$, $\xi^{(4)}$ can be imaginary or
complex numbers. Since the energy does not depend on the sign of $\xi$, we can
form a wave function from superposition of eight waves. From the boundary
conditions (\ref{eq:b-armchair}) resulting resulting set of linear equations
can have nonzero solution only if $8\times8$ determinant is zero. Analytical
form of this condition in is too large and too complicated to be useful.

For bilayer graphene zigzag nanotubes one has the periodic boundary condition
in the $y$ direction and the condition
\begin{equation}
\psi_{m,0,r_1}=\psi_{m,N+1,l_1}=\psi_{m,0,l_2}=\psi_{m,N+1,r_2}=0
\label{eq:b-zigzag}
\end{equation}
for the $x$ direction. Here $p=1,2$ is the number of the layer. This condition
is the same for both the AB-$\alpha$ and AB-$\beta$ stackings. In the bilayer
graphene there are two eigenstates with wave vectors along $x$ direction,
$\kappa^{(1)}$ and $\kappa^{(2)}$, having different absolute values but
corresponding the same energy: $E(\kappa^{(1)},\xi)=E(\kappa^{(2)},\xi)$. One
or both of the wave vectors $\kappa^{(1)}$ , $\kappa^{(2)}$ can be imaginary.
The energy can be equal only if the signs $s_1$, $s_2$ obey the condition
\begin{equation}
s_1^{(2)}s_2^{(2)}=-s_1^{(1)}s_2^{(1)}
\label{eq:cond-signs}
\end{equation}
When the bias potential is zero, $V=0$, from the equality of the energy we can
express $\kappa^{(2)}$:
\begin{equation}
s_3^{(2)}\cos\left(\frac{\kappa^{(2)}}{2}\right)=s_3^{(1)}\cos\left(\frac{
\kappa^{(1)}}{2}\right)+s_1^{(1)}s_2^{(1)}\frac{\gamma}{2\cos\left(\frac{\xi}{
2}\right)}E(\kappa^{(1)},\xi)
\label{eq:kappa2-kappa1}
\end{equation}
When $V\neq0$ then
\begin{equation}
s_3^{(2)}\cos\left(\frac{\kappa^{(2)}}{2}\right)=s_3^{(1)}\cos\left(\frac{
\kappa^{(1)}}{2}\right)\pm\frac{\gamma}{2\cos\left(\frac{\xi}{2}\right)}\sqrt{
4E^2V^2+\gamma^2(E^2-V^2)}
\end{equation}
There are four boundary conditions in the $x$ direction. Since the energy does
not depend on the sign of $\kappa$, we can form a wave function from
superposition of four waves. From the boundary conditions (\ref{eq:b-zigzag})
resulting set of linear equations can have nonzero solution only if $4\times4$
determinant is zero. The possible values of the wave vector $\xi$ is given by
Eq.~(\ref{eq:ksi-periodic}), and the conditions for the possible values of the
wave vector $\kappa$ are given in the Appendix \ref{appendixB}.

For $N\times\mathcal{N}$ sheet of bilayer graphene open boundary conditions in
the $y$ direction are the same as for armchair nanotubes,
Eq.~(\ref{eq:b-armchair}) and in the $x$ direction are the same as for zigzag
nanotubes, Eq.~(\ref{eq:b-zigzag}). For AB-$\alpha$ stacking, the conditions
for the possible values of the wave vectors $\kappa,\xi$ are combination of the
conditions for zigzag and armchair bilayer graphene tubes. Specifically, when
$V=0$, the conditions are given by Eqs.~(\ref{eq:cond-ksi}) and
(\ref{eq:cond-kappa-1}) or (\ref{eq:cond-kappa-2}). When $V\neq0$ then the
conditions are given by Eqs.~(\ref{eq:cond-ksi}) and (\ref{eq:cond-kappa-V}).
For AB-$\beta$ stacking it is impossible to separate conditions for the wave
vector $\xi$ from the conditions for the wave vector $\kappa$. The resulting
expressions are very large and complicated.

\subsection{Summary of the possible values of wave vectors}

For structures of bilayer graphene, the energy spectrum is completely
determined by Eq.~(\ref{eq:energy-bi}) or (\ref{eq:energy-bi-V}) with
appropriate expressions for wave vectors $\kappa$ and $\xi$.

Equations presented in Appendix \ref{appendixB} make one quantum number
dependent on the other. This dependence appears because of zigzag-shaped edges.
For structures where zigzag edges do not exist or their effect can be
disregarded the wave vector $\kappa_{\nu}$ can be replaced by a continuous
variable.

Thus, the possible values of wave vectors for various structures are as follows:
\begin{itemize}
\item For the armchair bilayer graphene ribbon of infinite length with
AB-$\alpha$ stacking, the wave vectors are determined by
\begin{equation}
0\leq\kappa\leq\pi\,,\qquad\xi_j=\frac{\pi j}{\mathcal{N}+1}\,,\qquad
j=1,\ldots,\mathcal{N}
\end{equation}

\item For the armchair bilayer graphene ribbon of infinite length with
AB-$\beta$ staking we have $0\leq\kappa\leq\pi$ and the equation for the
possible values of $\xi$ is complicated. \item For the zigzag bilayer graphene
ribbon of infinite length we have $0\leq\xi\leq\pi$, the conditions for the
possible values of $\kappa$, given in Appendix \ref{appendixB}, are different
for AB-$\alpha$ and AB-$\beta$ stackings. \item For the zigzag bilayer carbon
tube of infinite length with AB-$\alpha$ or AB-$\beta$ stacking, the wave
vectors are determined by
\begin{equation}
0\leq\kappa\leq\pi\,,\qquad\xi_j=\frac{2\pi}{\mathcal{N}}j\,,\qquad
j=-\left\lfloor\frac{\mathcal{N}}{2}\right\rfloor ,-\left\lfloor
\frac{\mathcal{N}}{2}\right\rfloor +1,\ldots,\left\lfloor
\frac{\mathcal{N}-1}{2}\right\rfloor
\end{equation}

\item For the armchair bilayer carbon tube of infinite length with AB-$\alpha$
or AB-$\beta$ stacking:
\begin{equation}
\kappa_{\nu}=\frac{2\pi}{N}\nu\,,\qquad\nu=-\left\lfloor
\frac{N}{2}\right\rfloor ,-\left\lfloor\frac{N}{2}\right\rfloor
+1,\ldots,\left\lfloor\frac{N-1}{2}\right\rfloor\,,\qquad0\leq\xi\leq\pi\,.
\end{equation}
Taking into account the ranges of the possible values of the wave vectors, zero
energy points for various structures with bias potential $V=0$ are as follows:
\item For zigzag bilayer carbon tube zero energy points are
$\left(0,\frac{2\pi}{3}\right)$, $\left(0,-\frac{2\pi}{3}\right)$. \item For
armchair bilayer carbon tube, zero energy point is
$\left(0,\frac{2\pi}{3}\right)$. \item The dispersion of armchair bilayer
graphene ribbon has only one zero-energy point $\left(0,\frac{2\pi}{3}\right)$
\item For zigzag bilayer graphene ribbon dispersion this point cannot be shown
in the real plane.
\end{itemize}

\section{Band structure near the Fermi energy}

\label{sec:fermi}In this Section only a part of the spectrum with smallest
absolute value of the energy is in focus. This part corresponds to $s_2=-1$,
$s_3=-1$. In order to obtain an approximate expression for the energy spectrum
near the Fermi energy we expand Eq.~(\ref{eq:phi2}) in power series near the
zero point $\kappa=0$, $\xi=2\pi/3$, yielding
\begin{eqnarray}
|\phi(\kappa,\xi)|^2 &\approx &
\frac{3}{4}\left[\frac{\kappa^2}{3}\left(1-\frac{\sqrt{3}}{2}q\right)
+q^2\left(1+\frac{q}{2\sqrt{3}}\right)\right]
\label{eq:fi2-approx}
\\ &\approx &
\frac{3}{4}\left(\frac{\kappa^2}{3}+\left(q-\frac{\kappa^2}{4\sqrt{
3}}\right)^2\right)\,.\nonumber
\end{eqnarray}
Here $q\equiv\xi-2\pi/3$ and $|\kappa|\ll1$, $|q|\ll1$. Substituting
Eq.~(\ref{eq:fi2-approx}) into Eq.~(\ref{eq:energy-bi}) or
Eq.~(\ref{eq:energy-bi-V}) one obtains the approximate expression for the
energy spectrum. Thus, when the bias potential is zero $V=0$, the approximate
expression for the energy is 
\begin{equation}
E(\kappa,\xi)=s_1\left(-\frac{\gamma}{2}+\sqrt{\frac{\gamma^2}{4}+\frac{3}{
4}\left(\frac{\kappa^2}{3}+\left(q-\frac{\kappa^2}{4\sqrt{
3}}\right)^2\right)}\right)
\end{equation}
Furthermore, assuming that $|\phi(\kappa,\xi)|^2\ll\gamma$, the branch of
Eq.~(\ref{eq:energy-bi}) with $s_2=-1$, $s_3=-1$ takes the form
\begin{equation}
E(\kappa,\xi)\approx
s_1\frac{|\phi(\kappa,\xi)|^2}{\gamma}
\label{eq:e-approx-bi}
\end{equation}
When $V\neq0$ and $V\ll\gamma$, then Eq.~(\ref{eq:energy-bi-V}) becomes
\begin{equation}
E(\kappa,\xi)\approx
s_1V-s_1\frac{2V}{\gamma^2}|\phi(\kappa,\xi)|^2+s_1\frac{|\phi(\kappa,\xi)|^4}{
2V\gamma^2}
\label{eq:e-approx-bi-V}
\end{equation}
The bilayer graphene has a gap at $|\phi(\kappa,\xi)|^2=2V^2$. However, since
the parameter $\gamma$ is small, $\gamma\ll1$, the approximate expressions
(\ref{eq:e-approx-bi}), (\ref{eq:e-approx-bi-V}) are suitable only for very
small values of $|\kappa|$ and $|q|$.

The spectrum of various structures of bilayer graphene can be obtained from the
approximate expressions for $|\phi(\kappa,\xi)|^2$ near zero points. In
contrast to Eqs.~(\ref{eq:e-approx-bi}) and (\ref{eq:e-approx-bi-V}), the
energy of monolayer graphene is $E(\kappa,\xi)=s_1\sqrt{|\phi(\kappa,\xi)|^2}$.
Thus, the analysis of the square root of $|\phi(\kappa,\xi)|^2$ essentially was
done in Ref.~\onlinecite{Onipko08}. Going back to the original wave vectors $k_x$ and
$k_y,$ the band structure of bilayer graphene tubes and ribbons when $V=0$
similarly as in Ref.~\onlinecite{Onipko08} can be summarized by the equation
\begin{equation}
E_{\nu}(k_{\Vert})\approx
s_1\left(-\frac{\gamma}{2}+\sqrt{\frac{\gamma^2}{4}+\frac{9}{4}a^2[(k_{\Vert}
-\bar{k}_{\Vert}^{\sigma})^2+k_{\bot\nu}^{\sigma2}]}\right)\,,
\label{eq:e-summary}
\end{equation}
where $k_{\Vert}$ and $k_{\bot\nu}$ denote the longitudinal (continuous) and
the transverse (quantized) components of the wave vector, respectively. Index
$\sigma$ specifies the structure. Further in this Section we will consider only
the case when $V=0$.

\subsection{Quantum conductance}

Within the framework of the Landauer approach\cite{Landauer57,*Landauer88,Buttiker86,Buttiker88},
the zero-temperature conductance of a ideal wire is equal to
\begin{equation}
G(E)=\frac{2e^2}{h}\sum_{\nu}g_{\nu}T_{\nu}(E)
\end{equation}
where $2e^2/h$ is conductance quantum, $g_{\nu}$ is the band degeneracy, and
transmission coefficient $T_{\nu}$ is zero or unity depending on whether the
$\nu$-th band is open or closed for charge carriers with energy $E$.

\begin{table}
\centering{}\caption{Degeneracy $g_{\nu}^{\sigma}$ of the $\nu$-th band energy
$|E_{\nu}^{\sigma}(k_{\Vert}=0)|$} \label{tbl:degeneracy}
\begin{tabular}{|c|c|}
\hline $\sigma$ & $g_{\nu}^{\sigma}$\tabularnewline \hline \hline armchair
bilayer carbon tube & $1$($\nu=0$), $2$($\nu\neq0$)\tabularnewline \hline
zigzag bilayer carbon tube & $2$\tabularnewline \hline armchair bilayer
graphene ribbon & $1$\tabularnewline \hline zigzag bilayer graphene ribbon &
$1$\tabularnewline \hline
\end{tabular}
\end{table}

When bias potential is zero, $V=0$, the transmission coefficient is
$T_{\nu}(E)=\Theta(E-E_{\nu}^{\sigma})$ for conduction bands and
$T_{\nu}(E)=\Theta(|E-E_{\nu}^{\sigma}|)$ for valence bands. Here
$E_{\nu}^{\sigma}$ are the subband threshold energies and $\Theta(x)$ is the
Heaviside step function. When the approximation Eq.~(\ref{eq:e-summary}) is
valid, the subband threshold energies are 
\begin{equation}
E_{\nu}^{\sigma}=s_1\left(-\frac{\gamma}{2}+\sqrt{\frac{\gamma^2}{4}+\frac{9}{
4}a^2k_{\bot\nu}^{\sigma2}}\right)\,.
\end{equation}
The degeneracies are shown in Table~\ref{tbl:degeneracy}. The values of
$g_{\nu}$ for armchair bilayer carbon tube and zigzag bilayer graphene ribbon
with $\nu>1$, represented in Table~\ref{tbl:degeneracy} should be doubled,
because electron or hole states with $\pm k_{\Vert}\neq0$ are degenerate.

The electron or hole conductance of armchair and zigzag bilayer carbon tubes
and their parent graphene ribbons has thus the form of a ladder, symmetrically
ascending with the increase in energy for electrons, and with the decrease of
energy for holes. For the charge carrier energy that falls between the $n$-th
and $(n+1)$-th bands, the wire conductance equals
\begin{equation}
G(E)=\frac{2e^2}{h}
\begin{cases}
n &\mbox{armchair bilayer ribbon}\\ 2n+2 &\mbox{zigzag bilayer ribbon}\\ 2n &
\mbox{zigzag bilayer carbon tube}\\ 2(2n+1) &\mbox{armchair bilayer carbon
tube}\end{cases}
\label{eq:conduct}
\end{equation}
The conductance for bilayer graphene ribbons has been numerically calculated in
Ref.~\onlinecite{Xu}. The expression (\ref{eq:conduct}) for the conductance coincides
with that of Ref.~\onlinecite{Xu}.

\subsection{Density of states}

The density of states (DOS) of a quantum wire, including a factor $2$ for the
spin degeneracy, reads
\begin{equation}
\rho(E)=\frac{2}{\pi}\sum_{\nu}\left(\frac{dE_{\nu}(k_{\Vert},k_{\bot\nu})}{dk_{
\Vert}}\right)^{-1}\,.
\end{equation}
The summation includes all transverse modes with energy $E_{\nu}\leq E$. Using
Eq.~(\ref{eq:e-summary}) we obtain the DOS of bilayer graphene
\begin{equation}
\rho(E)=\frac{2}{3\pi
a}\sum_{\nu}g_{\nu}\frac{(2|E|+\gamma)}{\sqrt{(|E|-|E_{\nu}^{\sigma}|)(|E|+|E_{
\nu}^{\sigma}|+\gamma)}}\Theta(|E|-|E_{\nu}^{\sigma}|)\,,
\label{eq:rho}
\end{equation}
The index $\nu$ is $\nu=0,\pm1,\pm2,\ldots$ for bilayer graphene tubes and
armchair bilayer graphene ribbons and $\nu=0,1,2,\ldots$ for the zigzag bilayer
graphene ribbons. The electron density at zero temperature is obtained by the
integration of the DOS from the charge neutrality point $\mu_0=0$ to the Fermi
energy,
\begin{equation}
n=\int_0^{E_F}\rho(E)dE
\end{equation}
Using Eq.~(\ref{eq:rho}) we get
\begin{equation}
n^{\sigma}(E_F)=\frac{4}{3\pi
a}\sum_{\nu}g_{\nu}\sqrt{(|E_F|-|E_{\nu}^{\sigma}|)(|E_F|+|E_{\nu}^{\sigma}|
+\gamma)}\Theta(|E_F|-|E_{\nu}|)
\end{equation}

\subsection{Armchair bilayer carbon tube}

\begin{figure}
\includegraphics[width=0.3\textwidth]{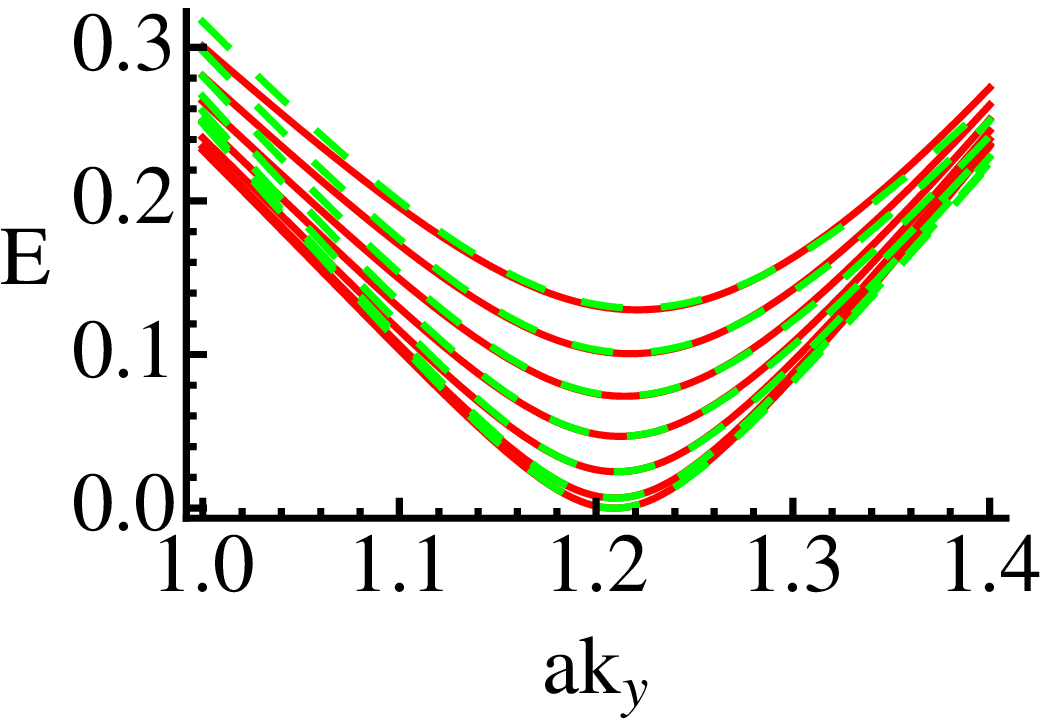}\includegraphics[width=0.3\textwidth]{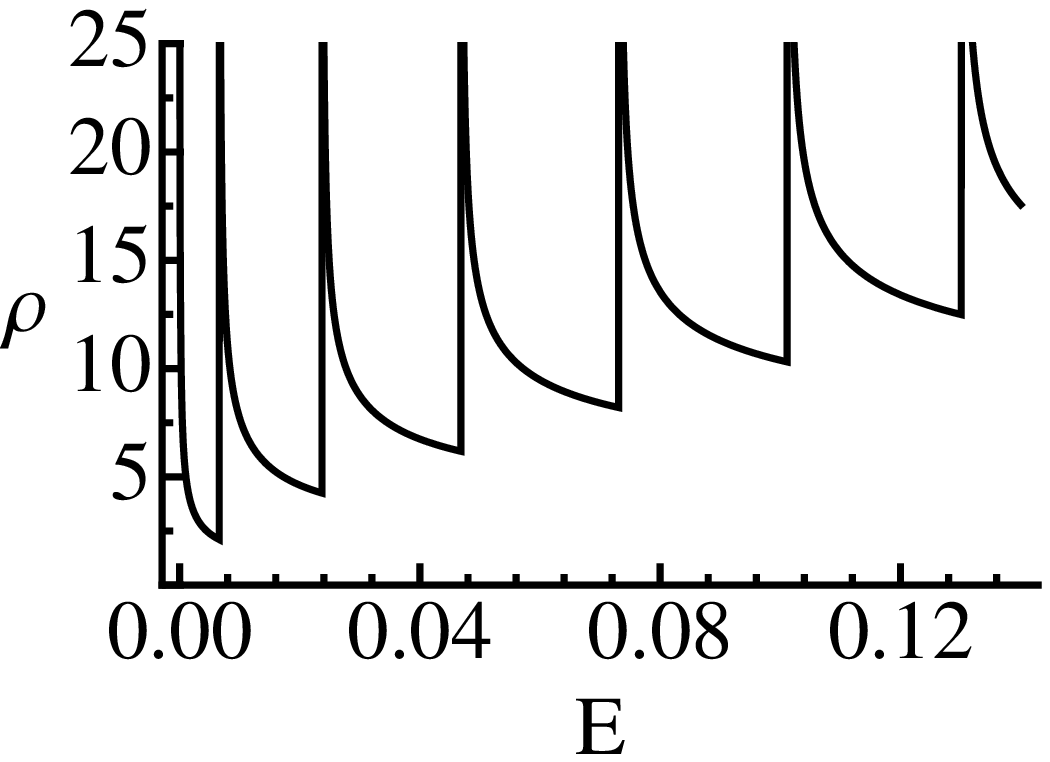}\includegraphics[width=0.3\textwidth]{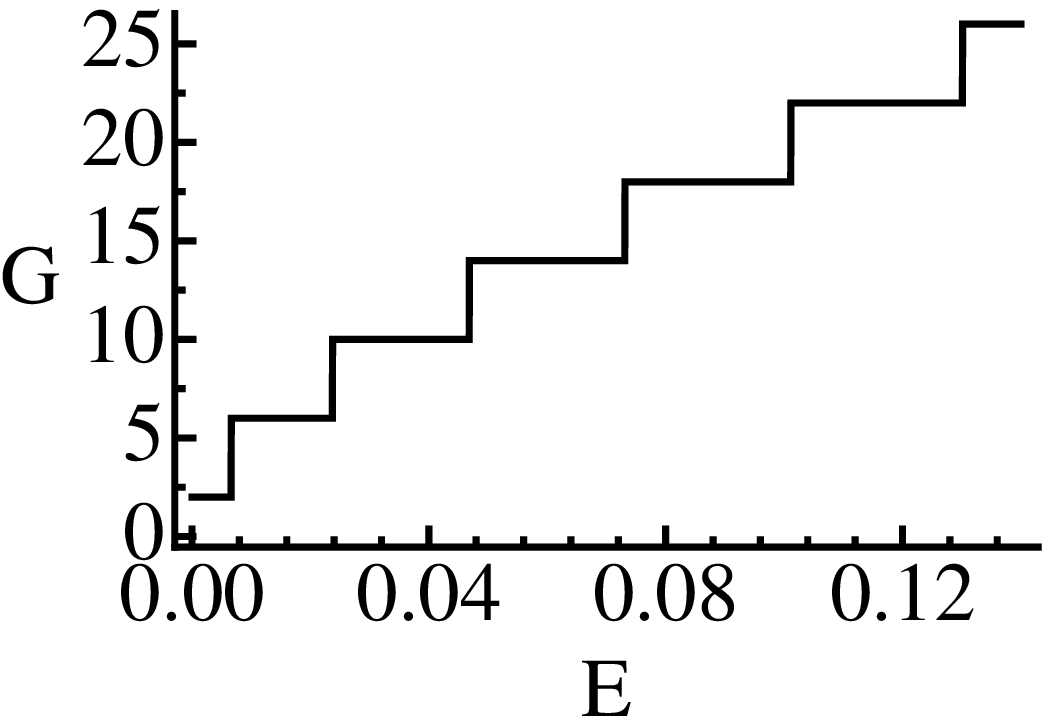}

\caption{(Color online) Band structure of armchair bilayer carbon tubes (left),
DOS (center) and conductance (right). The number of rectangular unit cells in
the $x$ direction $N=100$. Solid red lines are calculated according
Eqs.~(\ref{eq:energy-bi}), (\ref{eq:phi2}); dashed green lines represent
approximation (\ref{eq:energy-ABT}). Bands with $s_2=+1$ are not shown. DOS is
given in units of $a^{-1}$ and is calculated according to Eqs.~(\ref{eq:rho}),
(\ref{eq:threshold-ABT}). The conductance is given in units of $2e^2/h$ and is
calculated using Eq.~(\ref{eq:conduct}).}
\label{fig:ABT}
\end{figure}

For the armchair bilayer carbon tube we have that $\kappa$ in
Eq.~(\ref{eq:fi2-approx}) has discrete values $\kappa_{\nu}=2\pi\nu/N$,
$\nu=0,\pm1,\ldots$ and $q$ is continuous. Thus the energy spectrum has the form
\begin{equation}
E_{\nu}(k_y)=s_1\left(-\frac{\gamma}{2}+\sqrt{\frac{\gamma^2}{4}+\frac{9}{
4}a^2\left[\left(k_y-\bar{k}_{y,\nu}\right)^2+\frac{4\pi^2\nu^2}{
9a^2N^2}\right]}\right)
\label{eq:energy-ABT}
\end{equation}
with
\begin{equation}
\bar{k}_{y,\nu}=\frac{2\pi}{3\sqrt{3}a}+\frac{\pi^2\nu^2}{3aN^2}
\end{equation}
Conduction (valence) band bottoms (tops) are equal to
\begin{equation}
E_{\nu}=s_1\left(-\frac{\gamma}{2}+\sqrt{\frac{\gamma^2}{4}+\frac{\pi^2\nu^2}{
N^2}}\right)
\label{eq:threshold-ABT}
\end{equation}
The distance between the minima of the dispersion branches with the indices
$s_2=-1$ and $s_2=+1$ is $\gamma$. The number $\nu$ of subbands with the index
$s_2=-1$ and energy smaller than the energies of the subbands with index
$s_2=+1$ is greater than $1$ only when the number of rectangular unit cells in
the $x$ direction $N$ is sufficiently large. Using the equation
$E_{\nu,\mathrm{max}}=\gamma$ and estimating the subband threshold energy
(\ref{eq:threshold-ABT}) as $E_{\nu}\approx\pi^2\nu^2/(\gamma N^2)$ one obtains
that the requirement $\nu\gg1$ leads to $N\gg\pi/\gamma$. In calculations we
used $N=100$.

The band structure, calculated with the use of exact Eqs.~(\ref{eq:energy-bi}),
(\ref{eq:phi2}) and approximated according to Eq.~(\ref{eq:energy-ABT}), is
represented in Fig.~\ref{fig:ABT}. One sees that Eq.~(\ref{eq:energy-ABT})
provides accurate reproduction of exact results. Also is shown the DOS,
calculated with Eqs.~(\ref{eq:rho}), (\ref{eq:threshold-ABT}), and the
conductance $G(E)=(2e^2/h)2(2n+1)$.

\subsection{Zigzag bilayer carbon tube}

\begin{figure}
\includegraphics[width=0.3\textwidth]{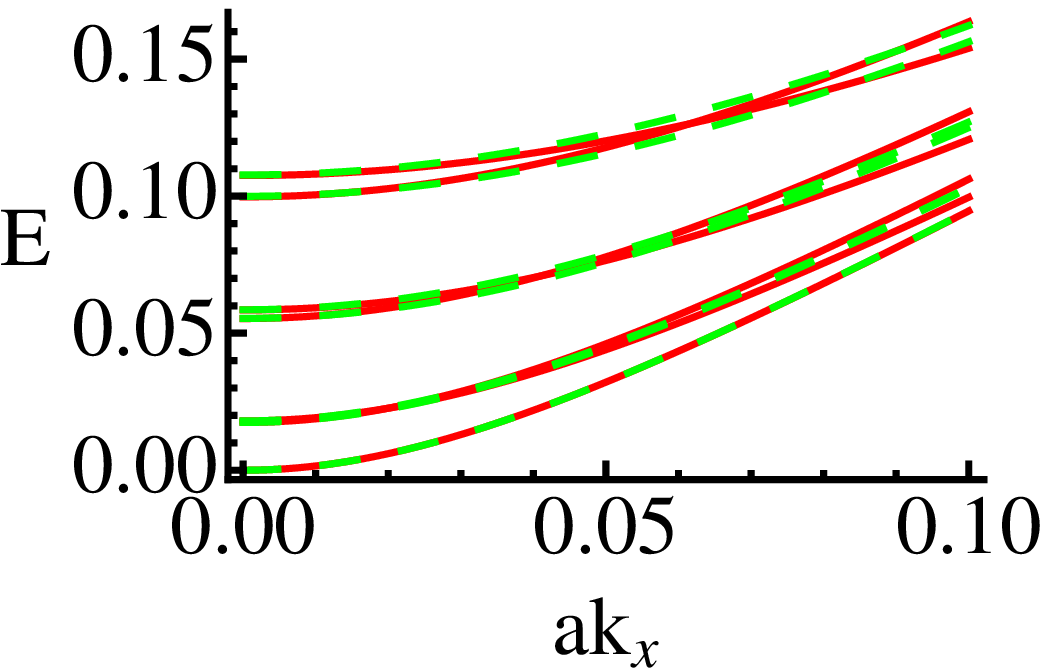}\includegraphics[width=0.3\textwidth]{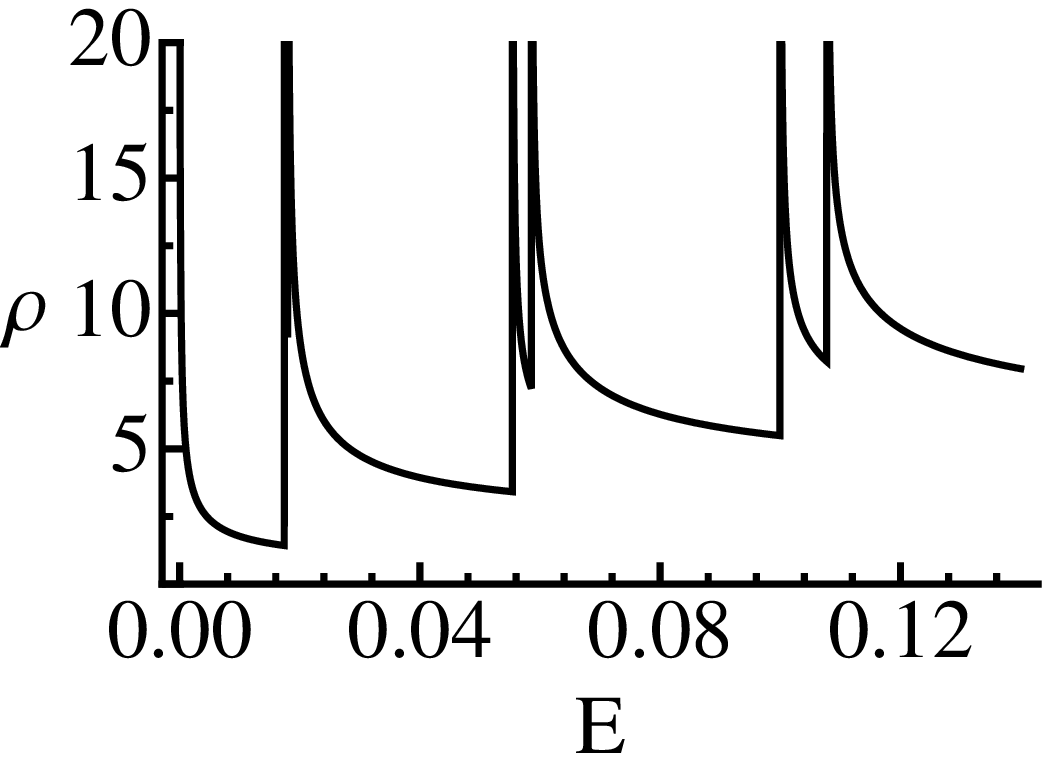}\includegraphics[width=0.3\textwidth]{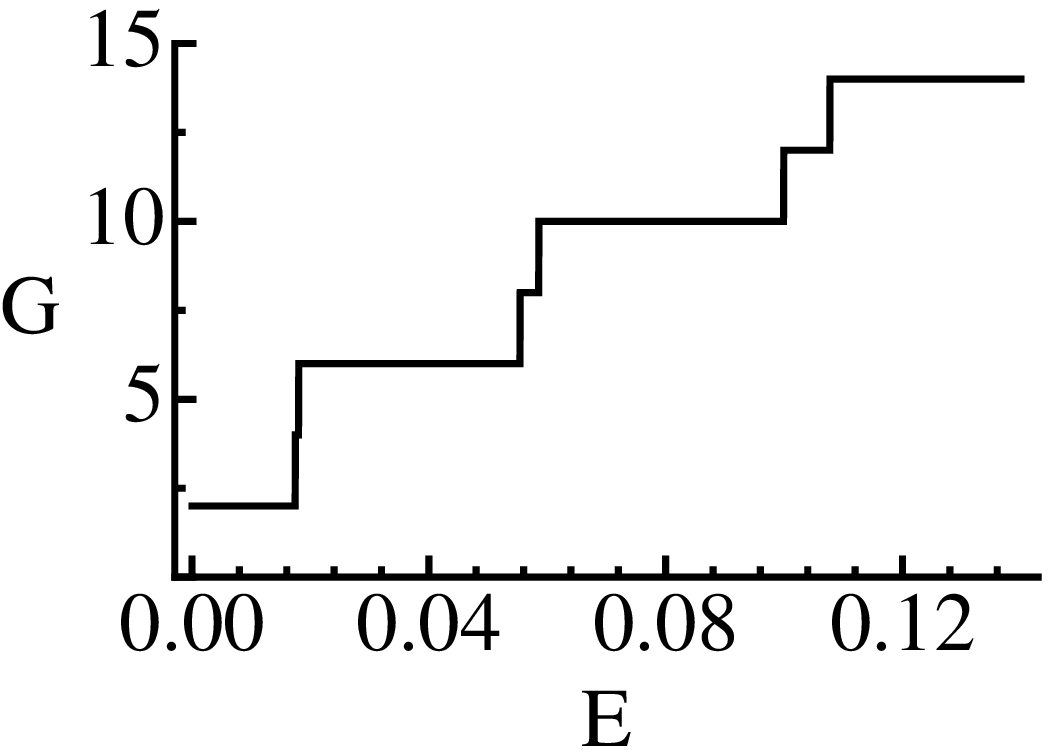}\newline
\includegraphics[width=0.3\textwidth]{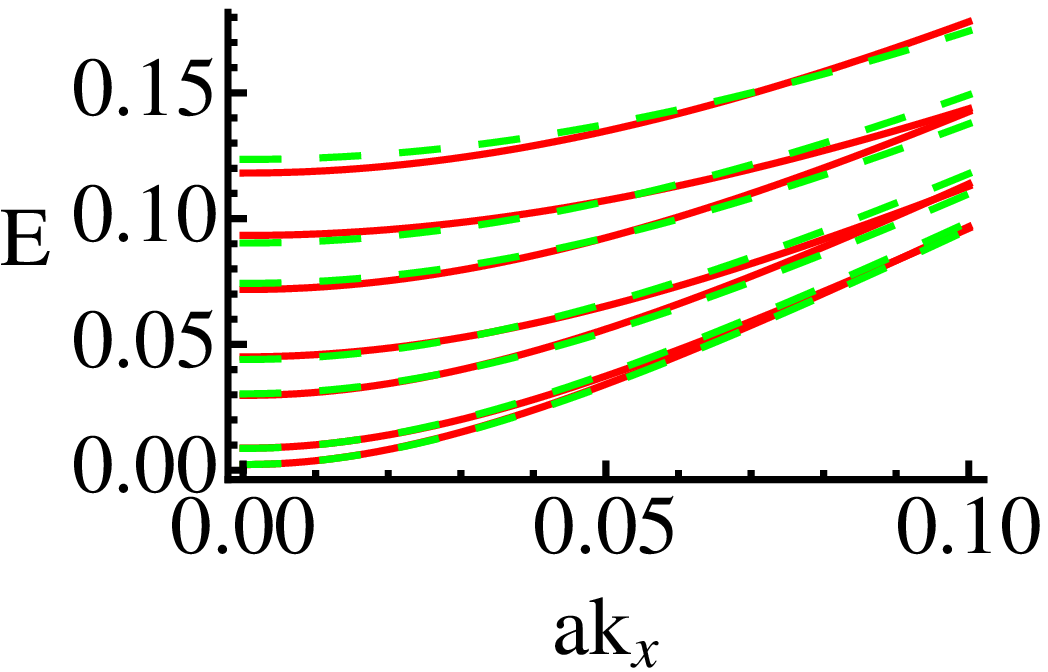}\includegraphics[width=0.3\textwidth]{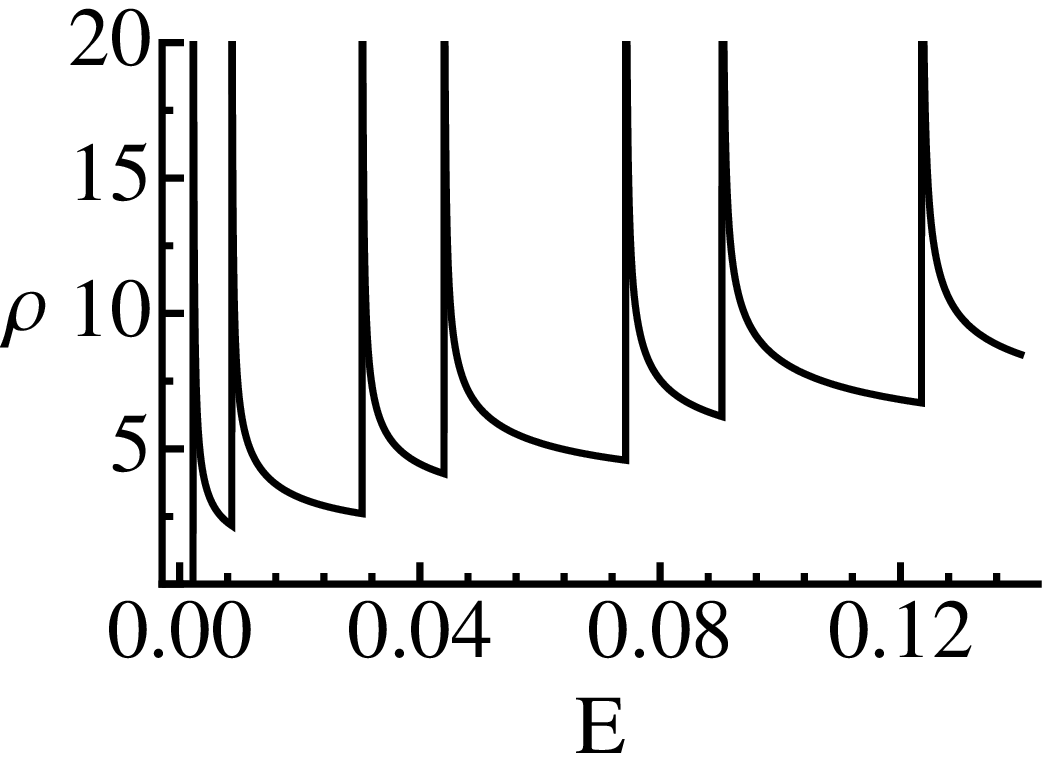}\includegraphics[width=0.3\textwidth]{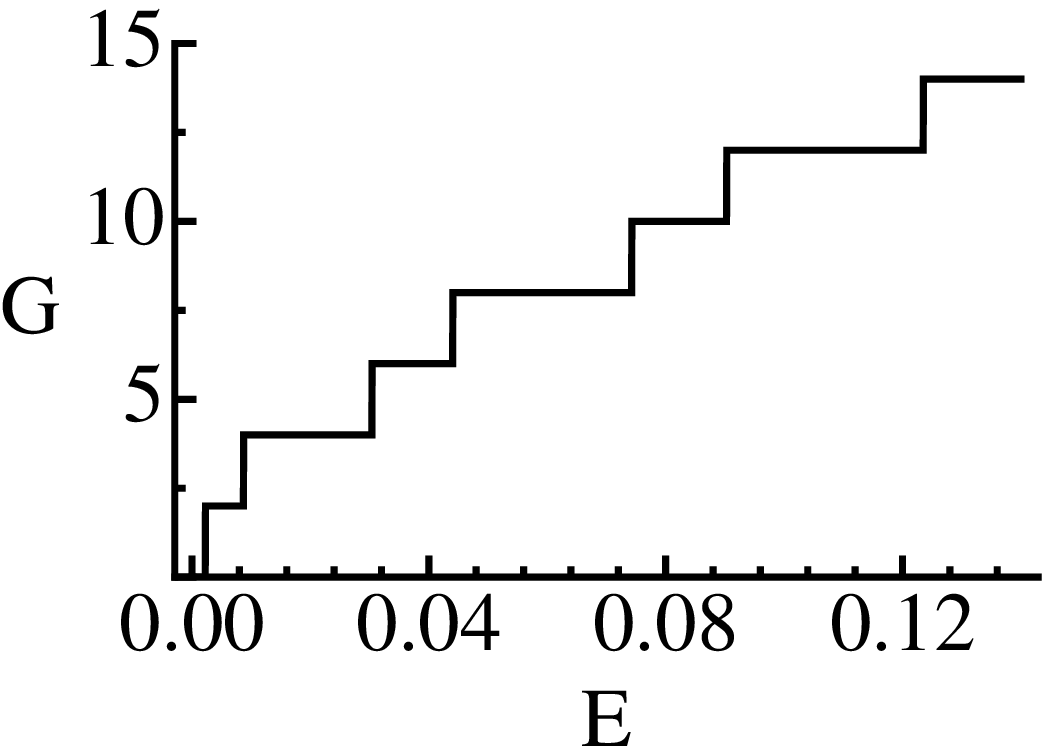}

\caption{(Color online) Upper part: band structure of metallic zigzag bilayer
carbon tubes (left), DOS (center) and conductance (right). The number of
hexagons in the $y$ direction $\mathcal{N}=102$. Lower part: band structure of
semiconducting zigzag bilayer carbon tubes (left), DOS (center) and conductance
(right). The number of hexagons in the $y$ direction $\mathcal{N}=100$. Solid
red lines are calculated according Eqs.~(\ref{eq:energy-bi}), (\ref{eq:phi2});
dashed green lines represent approximation (\ref{eq:energy-ZBT}). Bands with
$s_2=+1$ are not shown. DOS is given in units of $a^{-1}$ and is calculated
according to Eqs.~(\ref{eq:rho}), (\ref{eq:treshold-ZBT}). The conductance is
given in units of $2e^2/h$ and is calculated using Eq.~(\ref{eq:conduct}).}
\label{fig:ZBTm}
\end{figure}

Distinct from armchair bilayer carbon tubes, which are always metallic when
$V=0$, zigzag bilayer carbon tube has a gapless spectrum if
$j^*\equiv\mathcal{N}/3$ is an integer, $E_{j=j^*}(\kappa=0)=0$. Otherwise,
zigzag bilayer carbon tube spectrum has a gap. If $\mathcal{N}/3$ is not an
integer, band index of the lowest conduction (highest valence) band can be
equal either to $j^*\equiv(\mathcal{N}-1)/3$ or to
$j^*\equiv(\mathcal{N}+1)/3$. As a result of expansion near zero-energy points
in powers of $\kappa$ and $2\pi(j-j^*)/\mathcal{N}$, we arrive at
\begin{equation}
|\phi(\kappa,\xi)|^2\approx\frac{3}{4}\left(q_{\nu}^2+\frac{\kappa^2}{3}\right)
\label{eq:fi2-approx-2}
\end{equation}
where
\[
q_{\nu}=
\begin{cases}
\frac{2\pi}{\mathcal{N}}\left|\nu-\frac{1}{3}\right|\ll1\,, &
\mathrm{semiconducting}\\
\frac{2\pi|\nu|}{\mathcal{N}}\left(1+\frac{\pi\nu}{2\sqrt{3}\mathcal{
N}}\right)\ll1\,, &\mathrm{metallic}
\end{cases}
\]
with $\nu=0,\pm1,\ldots$ . The wave vector component $\kappa$ is continuous.
The energy spectrum has the form
\begin{equation}
E_{\nu}(k_x)=s_1\left(-\frac{\gamma}{2}+\sqrt{\frac{\gamma^2}{4}+\frac{9}{
4}a^2\left[k_x^2+\frac{q_{\nu}^2}{3a^2}\right]}\right)
\label{eq:energy-ZBT}
\end{equation}
Conduction (valence) band bottoms (tops) are equal to
\begin{equation}
E_{\nu}=s_1\left(-\frac{\gamma}{2}+\sqrt{\frac{\gamma^2}{4}+\frac{3}{4}q_{
\nu}^2}\right)
\label{eq:treshold-ZBT}
\end{equation}
The number $\nu$ of subbands with the index $s_2=-1$ and energy smaller than
the energies of the subbands with index $s_2=+1$ is greater than $1$ only when
the number of hexagons in the $y$ direction $\mathcal{N}$ is sufficiently
large. Approximating Eq.~(\ref{eq:treshold-ZBT}) as
$3\pi^2\nu^2/(\gamma\mathcal{N}^2)$ we get $\mathcal{N}\gg\sqrt{3}\pi/\gamma$.
In calculations we used $\mathcal{N}=102$ for metallic tubes and
$\mathcal{N}=100$ for semiconducting tubes.

The band structure, calculated with the use of exact Eqs.~(\ref{eq:energy-bi}),
(\ref{eq:phi2}) and approximated according to Eq.~(\ref{eq:energy-ZBT}) for
metallic and for semiconducting tubes is shown in Fig.~\ref{fig:ZBTm}. One sees
that Eq.~(\ref{eq:energy-ZBT}) provides accurate reproduction of exact results.
Also is shown the DOS, calculated with Eqs.~(\ref{eq:rho}),
(\ref{eq:treshold-ZBT}), and the conductance $G(E)=(2e^2/h)2n$.

\subsection{Armchair bilayer graphene ribbon}

\begin{figure}
\includegraphics[width=0.3\textwidth]{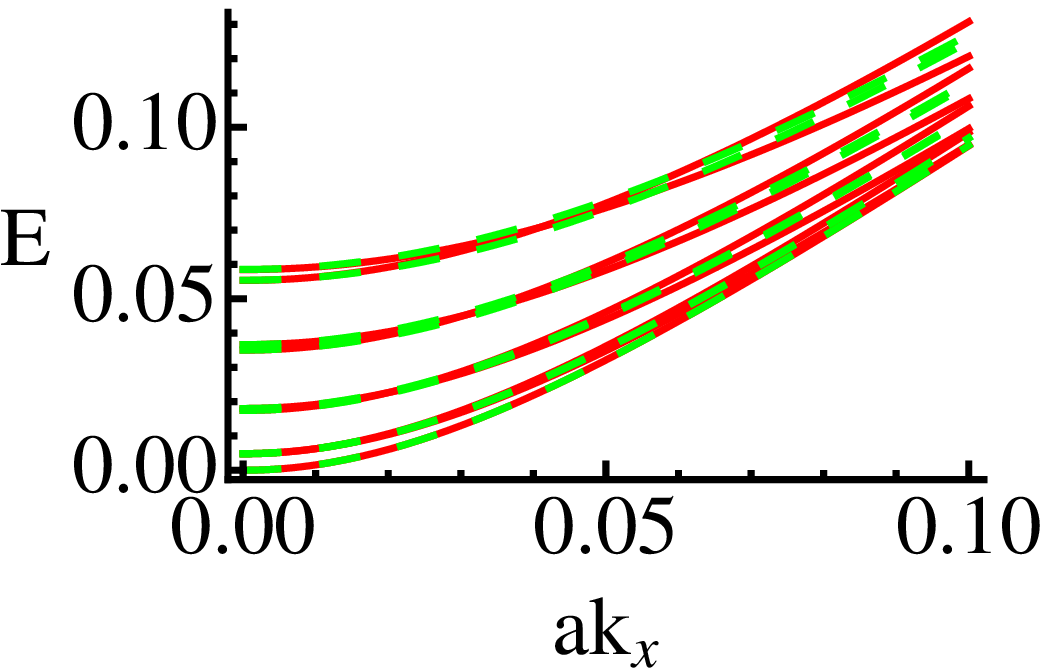}\includegraphics[width=0.3\textwidth]{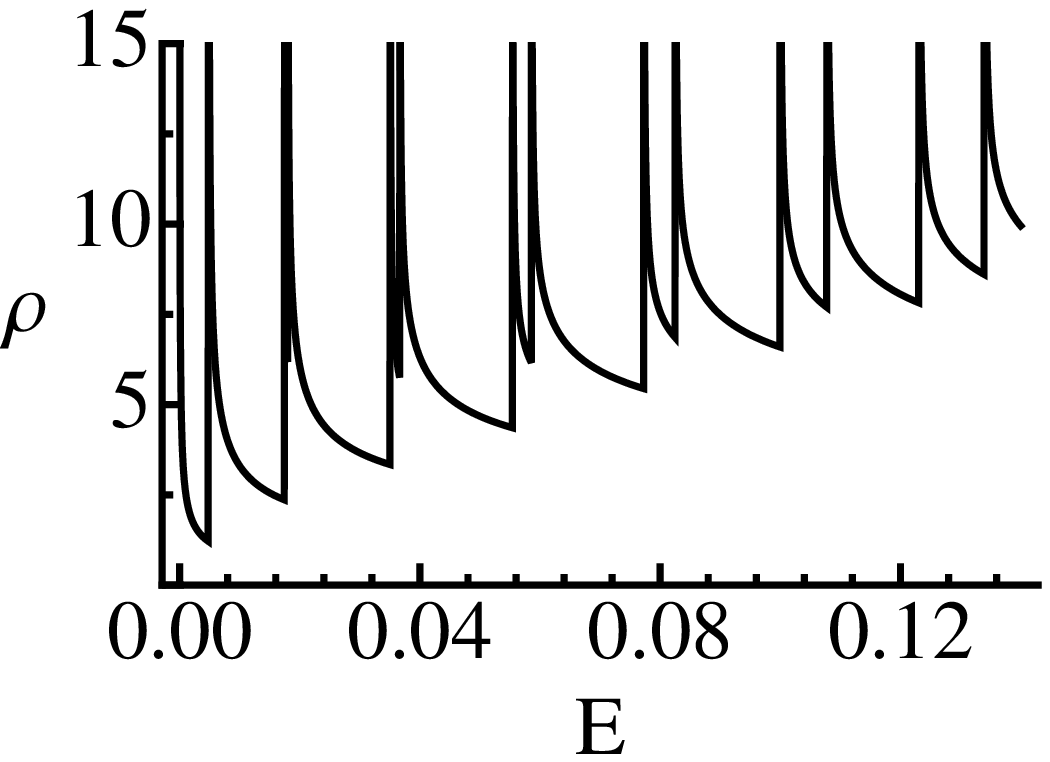}\includegraphics[width=0.3\textwidth]{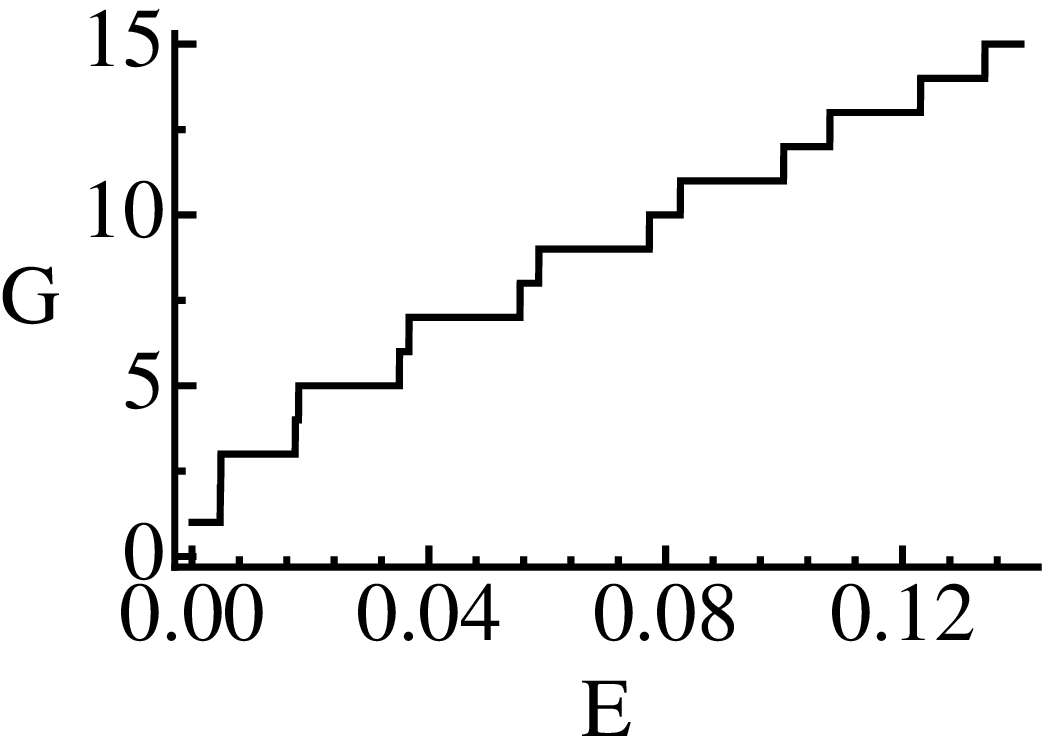}\newline
\includegraphics[width=0.3\textwidth]{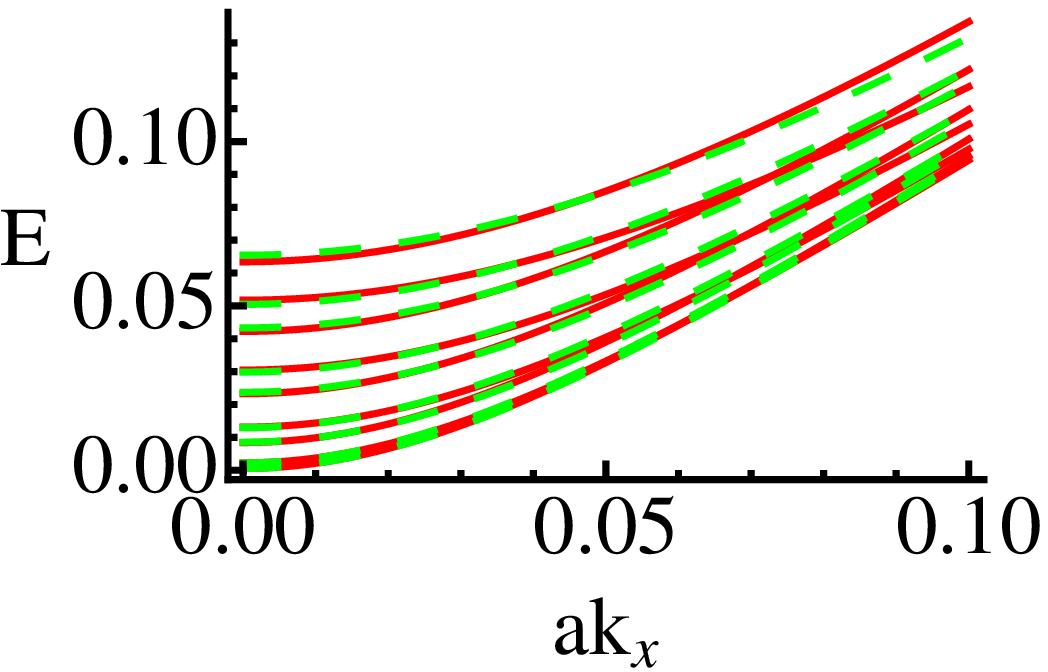}\includegraphics[width=0.3\textwidth]{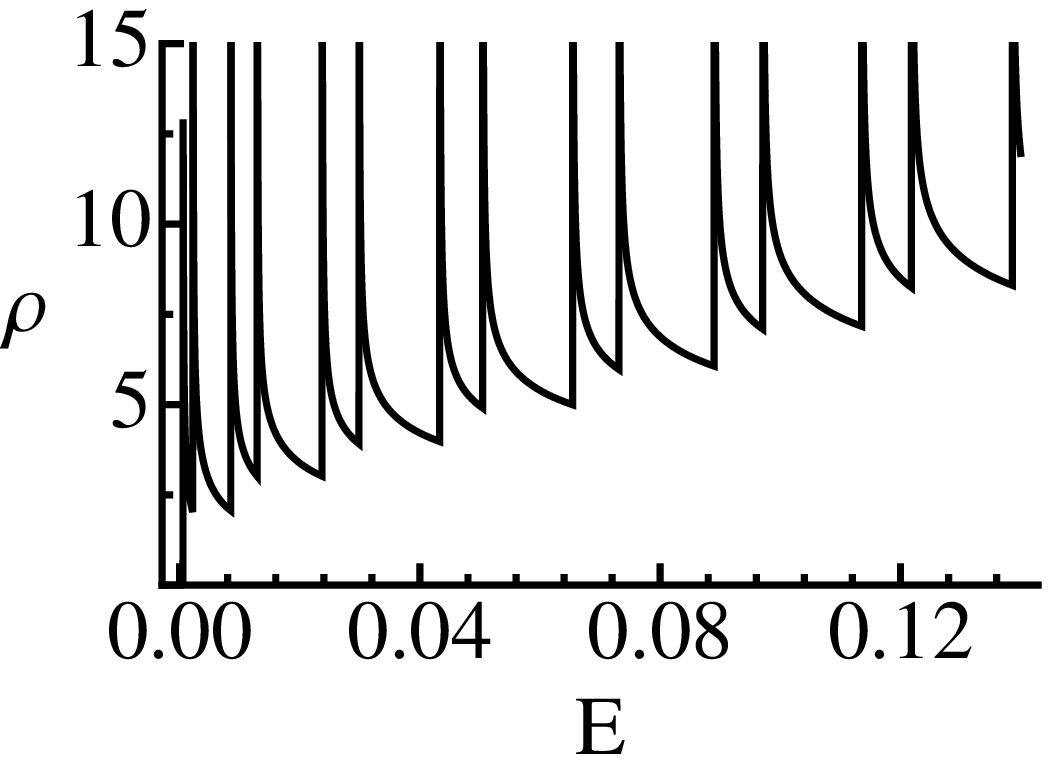}\includegraphics[width=0.3\textwidth]{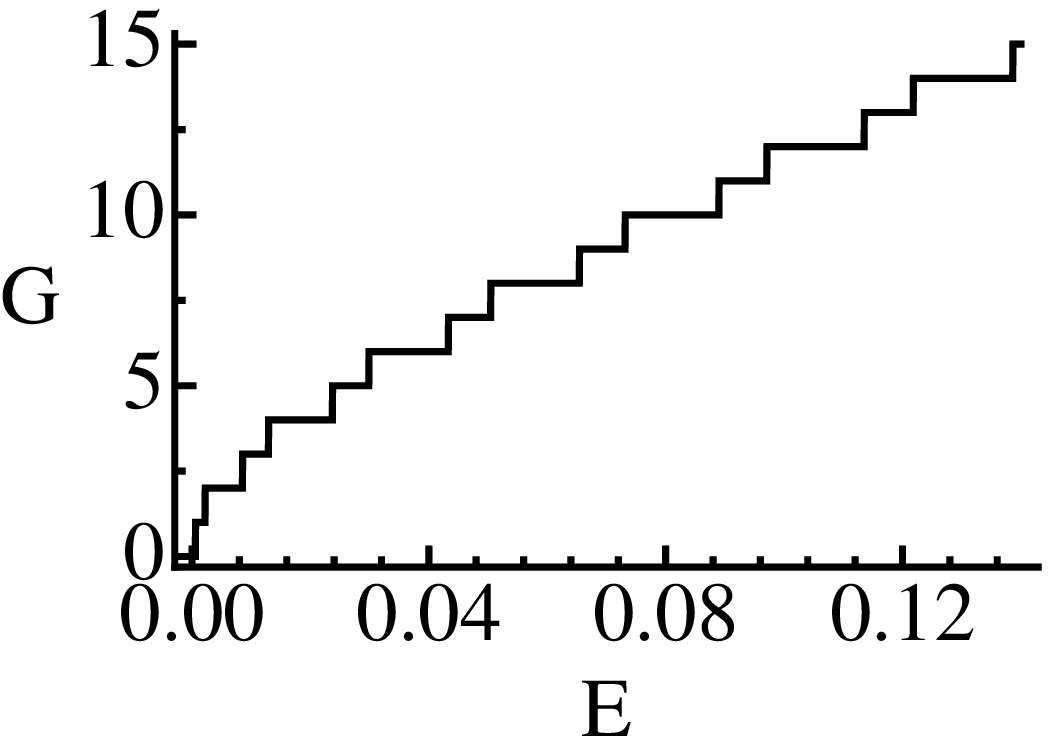}

\caption{(Color online) Upper part: band structure of metallic armchair bilayer
graphene ribbon with AB-$\alpha$ stacking (left), DOS (center) and conductance
(right). The number of hexagons in the $y$ direction $\mathcal{N}=101$. Lower
part: band structure of semiconducting armchair bilayer graphene ribbon with
AB-$\alpha$ stacking (left), DOS (center) and conductance (right). The number
of hexagons in the $y$ direction $\mathcal{N}=100$. Solid red lines are
calculated according Eqs.~(\ref{eq:energy-bi}), (\ref{eq:phi2}); dashed green
lines represent approximation (\ref{eq:energy-ABR}). Bands with $s_2=+1$ are
not shown. DOS is given in units of $a^{-1}$ and is calculated according to
Eqs.~(\ref{eq:rho}), (\ref{eq:treshold-ABR}). The conductance is given in units
of $2e^2/h$ and is calculated using Eq.~(\ref{eq:conduct}).}
\label{fig:ABRm}
\end{figure}

For the armchair bilayer graphene ribbon with AB-$\alpha$ stacking, the
condition for the wave-vector component $\xi$ has a simple expression. When
$V=0$ and $j^*\equiv2(\mathcal{N}+1)/3$ is an integer then the armchair bilayer
graphene ribbon is metallic. Then index $\nu=j-j^*=0$ corresponds to the
zero-energy band. If $2(\mathcal{N}+1)/3$ is not an integer, armchair bilayer
graphene ribbon spectrum has a gap, and the band closest to zero is either
$j^*\equiv(2\mathcal{N}+1)/3$ or $j^*\equiv(2\mathcal{N}+3)/3$ depending on
which of these two numbers is an integer. For $\kappa,\nu/\mathcal{N}\ll1$ we
get Eq.~(\ref{eq:fi2-approx-2}) with
\begin{equation}
q_{\nu}=
\begin{cases}
\frac{\pi}{\mathcal{N}+1}\left|\nu-\frac{1}{3}\right|\ll1\,, &
\mathrm{semiconducting}\\
\frac{\pi|\nu|}{\mathcal{N}+1}\left(1+\frac{\pi\nu}{4\sqrt{3}(\mathcal{N}
+1)}\right)\ll1\,, &\mathrm{metallic}\end{cases}
\end{equation}
with $\nu=0,\pm1,\ldots$ . The wave vector component $\kappa$ is continuous.
The difference between the boundary conditions for armchair bilayer graphene
ribbons and zigzag bilayer graphene tubes results in about two-times smaller
band spacing in the armchair ribbon spectrum than it was found for the zigzag
tube spectrum.

The energy spectrum has the form
\begin{equation}
E_{\nu}(k_x)=s_1\left(-\frac{\gamma}{2}+\sqrt{\frac{\gamma^2}{4}+\frac{9}{
4}a^2\left[k_x^2+\frac{q_{\nu}^2}{3a^2}\right]}\right)
\label{eq:energy-ABR}
\end{equation}
Conduction (valence) band bottoms (tops) are equal to
\begin{equation}
E_{\nu}=s_1\left(-\frac{\gamma}{2}+\sqrt{\frac{\gamma^2}{4}+\frac{3}{4}q_{
\nu}^2}\right)
\label{eq:treshold-ABR}
\end{equation}
The number $\nu$ of subbands with the index $s_2=-1$ and energy smaller than
the energies of the subbands with index $s_2=+1$ is greater than $1$ only when
the number of hexagons in the $y$ direction $\mathcal{N}$ is sufficiently
large. Approximating Eq.~(\ref{eq:treshold-ABR}) as
$3\pi^2\nu^2/(4\gamma(\mathcal{N}+1)^2)$ we get
$\mathcal{N}\gg\sqrt{3}\pi/(2\gamma)$. In calculations we used
$\mathcal{N}=101$ for metallic ribbons and $\mathcal{N}=100$ for semiconducting
ribbons.

The band structure, calculated with the use of exact Eqs.~(\ref{eq:energy-bi}),
(\ref{eq:phi2}) and approximated according to Eq.~(\ref{eq:energy-ABR}) for
metallic and semiconducting ribbons is shown in Fig.~\ref{fig:ABRm}. One sees
that Eq.~(\ref{eq:energy-ABR}) provides accurate reproduction of exact results.
Also is shown the DOS, calculated with Eqs.~(\ref{eq:rho}),
(\ref{eq:treshold-ABR}), and the conductance $G(E)=(2e^2/h)n$.

For the armchair bilayer graphene ribbon with AB-$\beta$ stacking there are no
explicit expressions for the possible values of $q$.

\subsection{Zigzag bilayer graphene ribbon}

\begin{figure}
\includegraphics[width=0.3\textwidth]{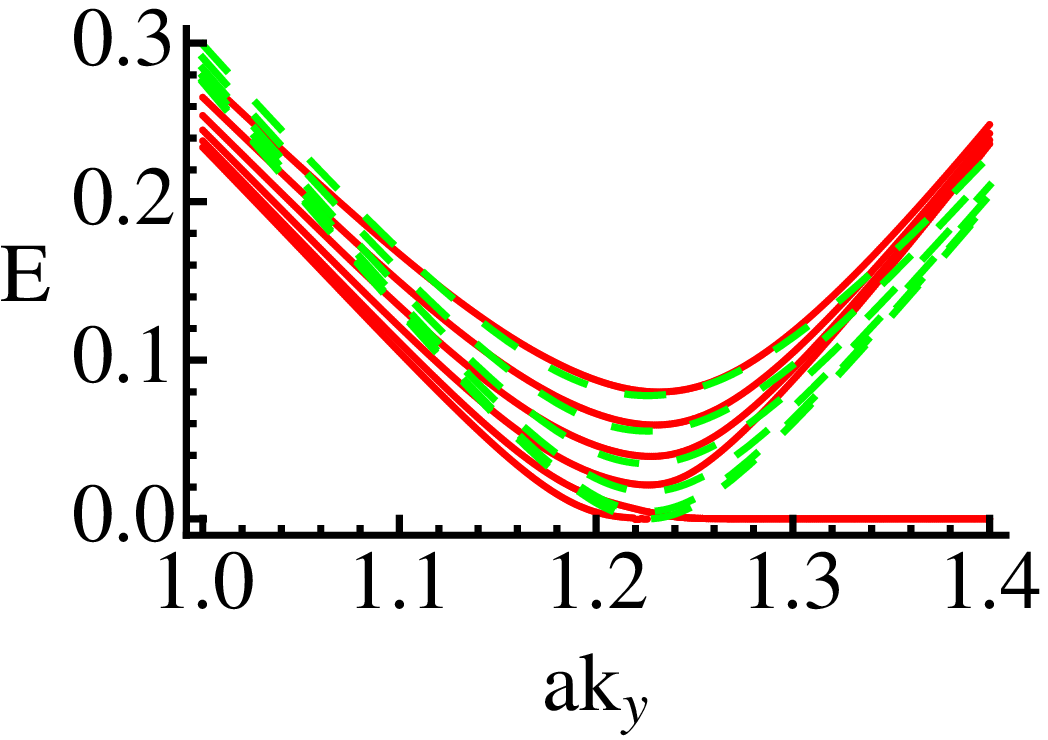}\includegraphics[width=0.3\textwidth]{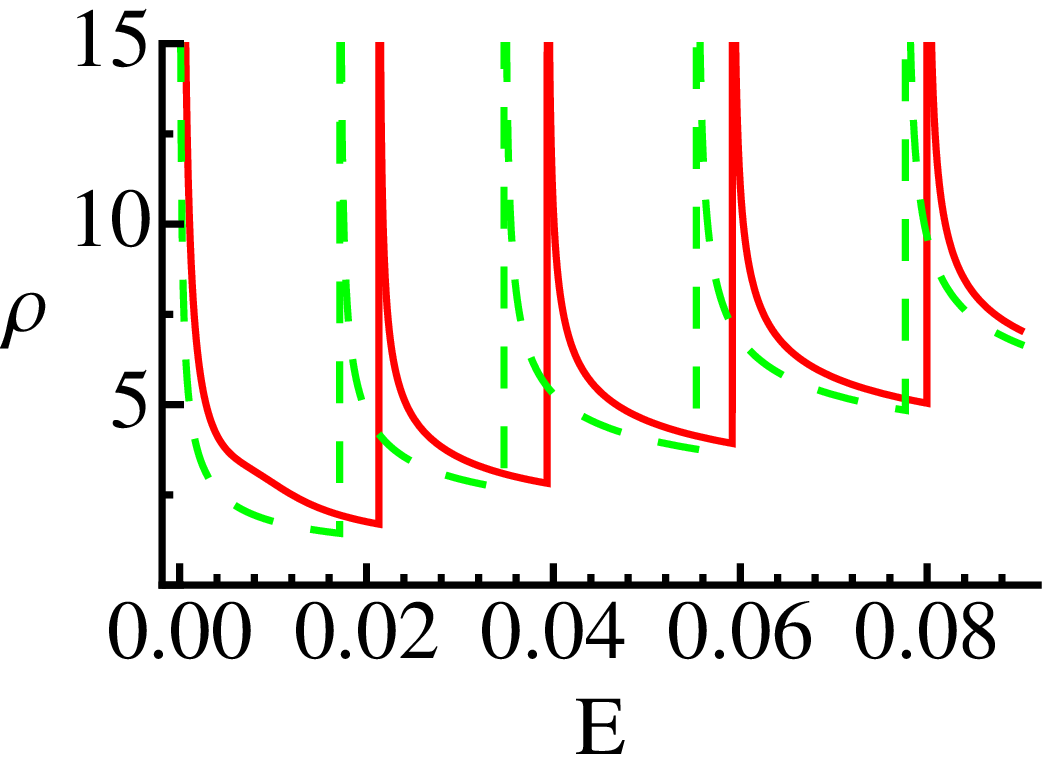}\includegraphics[width=0.3\textwidth]{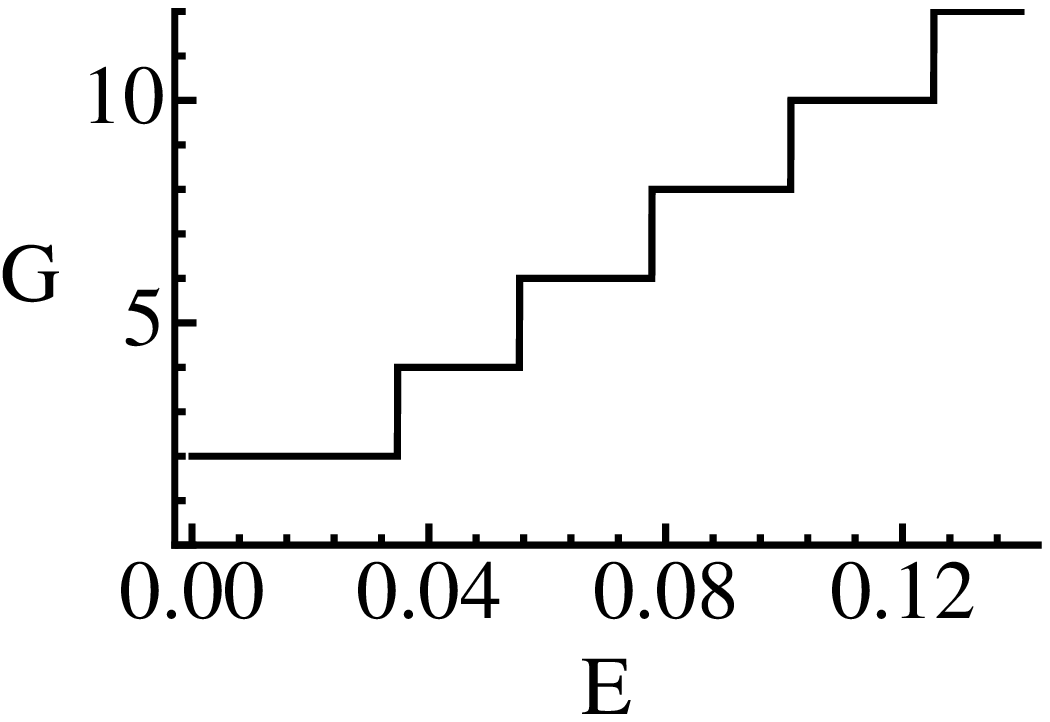}

\caption{(Color online) Band structure of zigzag bilayer graphene ribbon with
AB-$\alpha$ stacking (left), DOS (center) and conductance (right). The number
of rectangular unit cells in the $x$ direction $N=60$. Solid red lines are
calculated according Eqs.~(\ref{eq:energy-bi}), (\ref{eq:phi2}) with the
allowed values of the wave vector $\kappa$ obtained solving
Eqs.~(\ref{eq:cond-kappa-1}), (\ref{eq:cond-kappa-2}) and
(\ref{eq:kappa2-kappa1}); dashed green lines represent approximation
(\ref{eq:energy-ZBR}). Bands with $s_2=+1$ are not shown. DOS is given in units
of $a^{-1}$. The conductance is given in units of $2e^2/h$ and is calculated
using Eq.~(\ref{eq:conduct}).}
\label{fig:ZBR}
\end{figure}

In zigzag bilayer graphene ribbons the wave vector component $q$ is continuous
while the possible values of $\kappa$ are given by the solutions of the
equations (\ref{eq:kappa2-kappa1}) and (\ref{eq:cond-kappa-1}),
(\ref{eq:cond-kappa-2}) for AB-$\alpha$ stacking or
Eq.~(\ref{eq:cond-kappa-beta}) for AB-$\beta$ stacking, presented in Appendix
\ref{appendixB}. Since equations for $\kappa$ depend on the value of the wave
vector $\xi$, in zigzag bilayer graphene ribbons longitudinal and transverse
motions are not separable. When the energy of the subband with the index
$s_2=-1$ is smaller than the energies of the subbands with index $s_2=+1$, only
one of wave vectors $\kappa^{(1)}$ and $\kappa^{(2)}$ is real valued.
Therefore, the energy subbands can be labeled by the value of
$\kappa^{(1)}\equiv\kappa_{\nu}$ only.

Depending on the value of $|\xi|$, wave vectors $\kappa_{\nu}$ with $\nu=0,1$
can become imaginary. There are two critical values $\xi^{c(1)}$, $\xi^{c(2)}$
($\xi^{c(1)}<\xi^{c(2)}$) of the wave vector $\xi$, obtained by solving
Eqs.~(\ref{eq:kappa2-kappa1}) and (\ref{eq:cond-kappa-1}),
(\ref{eq:cond-kappa-2}) or (\ref{eq:cond-kappa-beta}) with $\kappa^{(1)}=0$.
When $\xi^{c(1)}<|\xi|<\xi^{c(2)}$ then one solution $\kappa_0$ becomes
imaginary whereas in the case $\xi^{c(2)}<|\xi|$ two solutions $\kappa_0$ and
$\kappa_1$ become imaginary. Both critical values $\xi^{c(1,2)}$ obey the
inequality $\xi^{c(1,2)}>2\pi/3$ and tend to the limit $2\pi/3$ as the number
$N$ grows. More tight lower bound of critical values is
$\xi^{c(1,2)}>2\arccos[N/(2N+1)]$. The imaginary solutions $\kappa_{\nu}$
represent edge states in zigzag bilayer graphene ribbons.

The energy bands of bilayer graphene are asymmetric near the point $q=0$. The
subbands (except corresponding to edge states) can be approximated by taking
$\kappa_{\nu}\approx\pi\nu/N$ and the minimum of of the subband located at
$\xi^{c(2)}$:
\begin{equation}
E_{\nu}(k_y)=s_1\left(-\frac{\gamma}{2}+\sqrt{\frac{\gamma^2}{4}+\frac{9}{
4}a^2\left[(k_y-\bar{k}_y)^2+\frac{\pi^2\nu^2}{9a^2N^2}\right]}\right)
\label{eq:energy-ZBR}
\end{equation}
with
\begin{equation}
\bar{k}_y=\frac{\xi^{c(2)}}{\sqrt{3}a}\,.
\end{equation}
The critical value $\xi^{c(2)}$ of the wave vector tends to the limit $2\pi/3$
as the number $N$ grows. Conduction (valence) band bottoms (tops) are equal to
\begin{equation}
E_{\nu}=s_1\left(-\frac{\gamma}{2}+\sqrt{\frac{\gamma^2}{4}+\frac{\pi^2\nu^2}{
4N^2}}\right)
\label{eq:treshold-ZBR}
\end{equation}
The number $\nu$ of of subbands with the index $s_2=-1$ and energy smaller than
the energies of the subbands with index $s_2=+1$ is greater than $1$ only when
the number of rectangular unit cells in the $x$ direction $N$ is sufficiently
large. Approximating Eq.~(\ref{eq:treshold-ZBR}) as $\pi^2\nu^2/(4\gamma N^2)$
we get $N\gg\pi/(2\gamma)$. In calculations we used $N=60$.

The band structure for zigzag bilayer graphene ribbons with AB-$\alpha$
stacking, calculated with the use of exact Eqs.~(\ref{eq:energy-bi}),
(\ref{eq:phi2}) with the allowed values of the wave vector $\kappa$ obtained
solving Eqs.~(\ref{eq:cond-kappa-1}), (\ref{eq:cond-kappa-2}) and
(\ref{eq:kappa2-kappa1}), as well as approximation (\ref{eq:energy-ZBR}) are
represented in Fig.~\ref{fig:ZBR}. Also is shown the DOS and the conductance
$G(E)=(2e^2/h)(2n+2)$. The DOS is calculated from exact band structure and also
using Eqs.~(\ref{eq:rho}), (\ref{eq:treshold-ABR}), taking the threshold
energies for $\nu=0,1$ to be $E_{\nu=0,1}=0$. The band structure of zigzag
bilayer graphene ribbons with AB-$\beta$ stacking is very similar to the band
structure of ribbons with AB-$\alpha$ stacking, only the critical values
$\xi^{c(1)}$, $\xi^{c(2)}$ are slightly different.

Taking the limit $N\rightarrow\infty$ in the zigzag bilayer graphene ribbon
with AB-$\beta$ stacking one can obtain the edge states of
Ref.~\onlinecite{Bilayer_edges}. However, care should be taken not to loose any solutions.
For large number $N$ we can write the absolute value of the imaginary wave
vector $i|\kappa|$ as $|\kappa|=\kappa^{(0)}+\delta$, where $\kappa^{(0)}$ is
the solution of the equation
\begin{equation}
e^{-\frac{\kappa^{(0)}}{2}}=2\cos(\xi/2)
\end{equation}
and $\delta$ is a small correction. From Eq.~(\ref{eq:energy-bi}) it follows
that such a value of $\kappa^{(0)}$ ensures the equality
$E(i\kappa^{(0)},\xi)=0$. Expanding Eq.~(\ref{eq:energy-bi}) in powers of
$\delta$ we get the approximate expression for the energy
\begin{equation}
E\approx\frac{s_1\delta}{\gamma}\left(2\cos^2\left(\frac{\xi}{2}\right)-\frac{
1}{2}\right)\,.
\label{eq:energy-corr}
\end{equation}
There are two eigenstates with wave vectors $\kappa^{(1)}$ and $\kappa^{(2)}$
having different absolute values but corresponding the same energy. From
Eqs.~(\ref{eq:energy-corr}) and (\ref{eq:cond-signs}) it follows that the
corrections to the wave vector obey the condition 
\begin{equation}
\delta^{(2)}=-\delta^{(1)}\,.
\label{eq:cond-corr}
\end{equation}
In Eq.~(\ref{eq:cond-kappa-beta}) taking into account only the first-order
terms with respect to $\delta$ one obtains the value of the correction
\begin{equation}
\delta=\pm2e^{-2\kappa^{(0)}N}(1-e^{-\kappa^{(0)}})\,.
\end{equation}
This expression for the correction is the same as for the single sheet of
graphene. The correction $\delta$ decreases exponentially with increasing the
number $N$.

For large $N$ it is sufficient to form the wave function obeying boundary
conditions (\ref{eq:b-zigzag}) as a superposition of two exponentially
decreasing terms with the wave vectors $i|\kappa^{(1)}|$ and $i|\kappa^{(2)}|$, 
\begin{equation}
\psi_{m,n,\alpha_p}=a^{(1)}c_{\alpha_p}(\xi_j,i|\kappa^{(1)}|)e^{i\xi_jm
-|\kappa^{(1)}|n}+a^{(2)}c_{\alpha_p}(\xi_j,i|\kappa^{(2)}|)e^{i\xi_jm-|\kappa^{
(2)}|n}\,.
\end{equation}
Substituting this expression for the wave function in the boundary conditions,
using Eqs.~(\ref{eq:c-ab-beta-1})--(\ref{eq:c-ab-beta-4}) and taking the limit
$N\rightarrow\infty$ we obtain two solutions for the coefficients $a^{(1)}$,
$a^{(2)}$: $a^{(2)}=0$ and $a^{(2)}=-a^{(1)}$. The wave function corresponding
to the solution $a^{(2)}=0$ is localized on the first layer, with the nonzero
coefficients $c_{l_1}$ and $c_{\rho_1}$. The wave function corresponding to the
solution $a^{(2)}=-a^{(1)}$ contains the difference
$e^{-|\kappa^{(1)}|n}-e^{-|\kappa^{(2)}|n}$. Expanding to the first order of
$\delta$ we get
\begin{equation}
e^{-|\kappa^{(1)}|n}-e^{-|\kappa^{(2)}|n}=e^{-(\kappa^{(0)}-\delta)n}-e^{
-(\kappa^{(0)}+\delta)n}\approx2\delta ne^{-\kappa^{(0)}n}\,.
\end{equation}
Taking the limit $N\rightarrow\infty$ and dropping the coefficients of the wave
function that are of the order of $\delta$ we obtain that nonzero coefficients
are $\psi_{m,n,l_1}$, $\psi_{m,n,\rho_1}$ in the first layer and
$\psi_{m,n,r_2}$, $\psi_{m,n,\lambda_2}$ in the second layer. The coefficients
$\psi_{m,n,r_2}$, $\psi_{m,n,\lambda_2}$ are proportional to
$e^{-\kappa^{(0)}n}$ while the coefficients $\psi_{m,n,l_1}$,
$\psi_{m,n,\rho_1}$ have $ne^{-\kappa^{(0)}n}$ behavior, as in
Ref.~\onlinecite{Bilayer_edges}.

\section{Conclusions}

\textbf{\label{sec:concl}}An exact analytical description of $\pi$ electron
spectrum based on tight-binding model of bilayer graphene has been presented.
The bilayer graphene structures considered in this article have rectangular
geometry and finite size in one or both directions with armchair- and
zigzag-shaped edges. This includes bilayer graphene nanoribbons and nanotubes.
The exact solution of the Schr\"odinger problem, the spectrum and wave
functions, has been obtained and used to analyze the density of states and the
conductance quantization. Our method brings a connection between $\pi$ electron
spectrum in infinite and finite-size bilayer graphene.
\begin{acknowledgments}
The authors acknowledge a collaborative grant from the Swedish Institute and a
grant No.\ MIP-123/2010 by the Research Council of Lithuania. I.V.Z
acknowledges a support from the Swedish Research Council (VR).
\end{acknowledgments}
\appendix

\section{Eigenvectors of bilayer graphene using rectangular unit cells}

\label{appendixA}The expressions for the coefficients of the eigenvectors are:
For AB-$\alpha$ stacking, $V=0$
\begin{eqnarray}
c_{r_1} & = & 1\,,\quad
c_{\rho_1}=-e^{-i\frac{\xi}{2}}\frac{E(\kappa,\xi)}{\phi(-\kappa,\xi)}\,,\\
c_{l_1} & = &
-s_3e^{-i\frac{\kappa}{2}}\frac{E(\kappa,\xi)}{\phi(-\kappa,\xi)}\,,\quad
c_{\lambda_1}=s_3e^{-i\frac{1}{2}(\kappa+\xi)}\,,\\ c_{r_2} & = &
-s_1s_2\frac{\phi(\kappa,\xi)}{\phi(-\kappa,\xi)}\,,\quad
c_{\rho_2}=s_1s_2e^{-i\frac{\xi}{2}}\frac{E(\kappa,\xi)}{\phi(
-\kappa,\xi)}\,,\\ c_{l_2} & = &
s_1s_2s_3e^{i\frac{\kappa}{2}}\frac{E(\kappa,\xi)}{\phi(-\kappa,\xi)}\,,\quad
c_{\lambda_2}=-s_1s_2s_3e^{i\frac{1}{2}(\kappa-\xi)}\frac{\phi(\kappa,\xi)}{
\phi(-\kappa,\xi)}\,.
\end{eqnarray}
For AB-$\alpha$ stacking, $V\neq0$
\begin{eqnarray}
c_{r_1} & = & 1\,,\quad
c_{\rho_1}=-e^{-i\frac{\xi}{2}}\frac{E+V}{\phi(-\kappa,\xi)}\,,\\ c_{l_1} &
= & -s_3e^{-i\frac{\kappa}{2}}\frac{E+V}{\phi(-\kappa,\xi)}\,,\quad
c_{\lambda_1}=s_3e^{-i\frac{1}{2}(\kappa+\xi)}\,,\\ c_{r_2} & = &
-\frac{\phi(\kappa,\xi)}{\phi(-\kappa,\xi)}f(\kappa,\xi)\,,\quad
c_{\rho_2}=s_1s_2e^{-i\frac{\xi}{2}}\frac{E-V}{\phi(
-\kappa,\xi)}f(\kappa,\xi)\,,\\ c_{l_2} & = &
s_1e^{i\frac{\kappa}{2}}\frac{E-V}{\phi(-\kappa,\xi)}f(\kappa,\xi)\,,\quad
c_{\lambda_2}=-s_3e^{i\frac{1}{2}(\kappa-\xi)}\frac{\phi(\kappa,\xi)}{\phi(
-\kappa,\xi)}f(\kappa,\xi)\,.
\end{eqnarray}
For AB-$\beta$ stacking, $V=0$
\begin{eqnarray}
c_{r_1} & = & 1\,,\quad
c_{\rho_1}=-e^{-i\frac{\xi}{2}}\frac{\phi(\kappa,\xi)}{E(\kappa,\xi)}\,,
\label{eq:c-ab-beta-1}
\\ c_{l_1} & = &
-s_3e^{-i\frac{\kappa}{2}}\frac{\phi(\kappa,\xi)}{E(\kappa,\xi)}\,,\quad
c_{\lambda_1}=s_3e^{-i\frac{1}{2}(\kappa+\xi)}\,,\\ c_{r_2} & = &
-s_1s_2s_3e^{i\frac{1}{2}(\xi-\kappa)}\,,\quad
c_{\rho_2}=s_1s_2s_3e^{-i\frac{\kappa}{2}}\frac{\phi(-\kappa,\xi)}{
E(\kappa,\xi)}\,,\\ c_{l_2} & = &
s_1s_2e^{i\frac{\xi}{2}}\frac{\phi(-\kappa,\xi)}{E(\kappa,\xi)}\,,\quad
c_{\lambda_2}=-s_1s_2\,.
\label{eq:c-ab-beta-4}
\end{eqnarray}
For AB-$\beta$ stacking, $V\neq0$
\begin{eqnarray}
c_{r_1} & = & 1\,,\quad
c_{\rho_1}=-e^{-i\frac{\xi}{2}}\frac{\phi(\kappa,\xi)}{E+V}\,,\\ c_{l_1} &
= & -s_3e^{-i\frac{\kappa}{2}}\frac{\phi(\kappa,\xi)}{E+V}\,,\quad
c_{\lambda_1}=s_3e^{-i\frac{1}{2}(\kappa+\xi)}\,,\\ c_{r_2} & = &
-s_3e^{i\frac{1}{2}(\xi-\kappa)}f(\kappa,\xi)\,,\quad
c_{\rho_2}=s_3e^{-i\frac{\kappa}{2}}\frac{\phi(-\kappa,\xi)}{E
-V}f(\kappa,\xi)\,,\\ c_{l_2} & = &
s_1s_2e^{i\frac{\xi}{2}}\frac{\phi(-\kappa,\xi)}{E-V}f(\kappa,\xi)\,,\quad
c_{\lambda_2}=-f(\kappa,\xi)\,.
\end{eqnarray}

\section{Wave vectors of zigzag bilayer carbon tubes}

\label{appendixB}For AB-$\alpha$ stacking and $V=0$ the possible values of
$\kappa_{j,\nu_j}^{(1)}$ are solutions of one of the equations
\begin{equation}
1+\cos\left(\frac{\xi}{2}\right)\left(s_3^{(1)}\frac{\cos\left(\frac{1}{
2}\kappa^{(1)}(N+1)\right)}{\cos\left(\frac{1}{2}\kappa^{(1)}N\right)}+s_3^{
(2)}\frac{\sin\left(\frac{1}{2}\kappa^{(2)}(N+1)\right)}{\sin\left(\frac{1}{
2}\kappa^{(2)}N\right)}\right)=0
\label{eq:cond-kappa-1}
\end{equation}
or
\begin{equation}
1+\cos\left(\frac{\xi}{2}\right)\left(s_3^{(1)}\frac{\sin\left(\frac{1}{
2}\kappa^{(1)}(N+1)\right)}{\sin\left(\frac{1}{2}\kappa^{(1)}N\right)}+s_3^{
(2)}\frac{\cos\left(\frac{1}{2}\kappa^{(2)}(N+1)\right)}{\cos\left(\frac{1}{
2}\kappa^{(2)}N\right)}\right)=0
\label{eq:cond-kappa-2}
\end{equation}
When $V\neq0$ then the equation for $\kappa$ reads
\begin{multline}
\frac{1}{4}(f(\kappa^{(1)})-f(\kappa^{(2)}))^{2}\left(1+2s_{3}^{(1)}\cos\left(\frac{\xi}{2}\right)\frac{\sin\left(\kappa^{(1)}\left(N+\frac{1}{2}\right)\right)}{\sin(\kappa^{(1)}N)}\right)\left(1+2s_{3}^{(2)}\cos\left(\frac{\xi}{2}\right)\frac{\sin\left(\kappa^{(2)}\left(N+\frac{1}{2}\right)\right)}{\sin(\kappa^{(2)}N)}\right)\\
-f(\kappa^{(1)})f(\kappa^{(2)})\cos^{2}\left(\frac{\xi}{2}\right)\left(s_{3}^{(1)}\frac{\cos\left(\frac{1}{2}\kappa^{(1)}(N+1)\right)}{\cos\left(\frac{1}{2}\kappa^{(1)}N\right)}-s_{3}^{(2)}\frac{\cos\left(\frac{1}{2}\kappa^{(2)}(N+1)\right)}{\cos\left(\frac{1}{2}\kappa^{(2)}N\right)}\right)\\
\times\left(s_{3}^{(1)}\frac{\sin\left(\frac{1}{2}\kappa^{(1)}(N+1)\right)}{\sin\left(\frac{1}{2}\kappa^{(1)}N\right)}-s_{3}^{(2)}\frac{\sin\left(\frac{1}{2}\kappa^{(2)}(N+1)\right)}{\sin\left(\frac{1}{2}\kappa^{(2)}N\right)}\right)=0\label{eq:cond-kappa-V}
\end{multline}
Here the function 
\begin{equation}
f(\kappa,\xi)=\frac{(E+V)^2-|\phi(\kappa,\xi)|^2}{\gamma(E-V)}
\end{equation}
describes the contribution of the second sheet of graphene to the eigenvector.

For AB-$\beta$ stacking and $V=0$ the possible values of $\kappa$ are solutions
of the equation
\begin{multline}
\left(1+2s_{3}^{(1)}\cos\left(\frac{\xi}{2}\right)\frac{\sin\left(\kappa^{(1)}\left(N+\frac{1}{2}\right)\right)}{\sin(\kappa^{(1)}N)}\right)\left(1+2s_{3}^{(2)}\cos\left(\frac{\xi}{2}\right)\frac{\sin\left(\kappa^{(2)}\left(N+\frac{1}{2}\right)\right)}{\sin(\kappa^{(2)}N)}\right)\\
+\frac{1}{2}s_{3}^{(1)}s_{3}^{(2)}\left(\cos\left(\frac{\kappa^{(1)}}{2}\right)\cos\left(\frac{\kappa^{(2)}}{2}\right)+\frac{1-\cos(\kappa^{(1)}N)\cos(\kappa^{(2)}N)}{\sin(\kappa^{(1)}N)\sin(\kappa^{(2)}N)}\sin\left(\frac{\kappa^{(1)}}{2}\right)\sin\left(\frac{\kappa^{(2)}}{2}\right)\right)-\frac{1}{2}=0\label{eq:cond-kappa-beta}
\end{multline}
When $V\neq0$ then the equation for $\kappa$ is
\begin{multline}
\frac{1}{4}(f(\kappa^{(1)})-f(\kappa^{(2)}))^{2}\left(1+2s_{3}^{(1)}\cos\left(\frac{\xi}{2}\right)\frac{\sin\left(\kappa^{(1)}\left(N+\frac{1}{2}\right)\right)}{\sin(\kappa^{(1)}N)}\right)\left(1+2s_{3}^{(2)}\cos\left(\frac{\xi}{2}\right)\frac{\sin\left(\kappa^{(2)}\left(N+\frac{1}{2}\right)\right)}{\sin(\kappa^{(2)}N)}\right)\\
+\frac{1}{2}f(\kappa^{(1)})f(\kappa^{(2)})\left(1-s_{3}^{(1)}s_{3}^{(2)}\left(\cos\left(\frac{\kappa^{(1)}}{2}\right)\cos\left(\frac{\kappa^{(2)}}{2}\right)\right.\right.\\
\left.\left.+\frac{1-\cos(\kappa^{(1)}N)\cos(\kappa^{(2)}N)}{\sin(\kappa^{(1)}N)\sin(\kappa^{(2)}N)}\sin\left(\frac{\kappa^{(1)}}{2}\right)\sin\left(\frac{\kappa^{(2)}}{2}\right)\right)\right)=0
\end{multline}

\end{document}